\documentclass[12pt,letterpaper]{JHEP3}
\usepackage{amsmath}
\usepackage{cite}
\usepackage{epsfig}
\usepackage[vcentermath]{youngtab}
\usepackage{epic}
\usepackage{supertabular}

\newdimen\tableauside\tableauside=1.0ex
\newdimen\tableaurule\tableaurule=0.4pt
\newdimen\tableaustep
\def\phantomhrule#1{\hbox{\vbox to0pt{\hrule height\tableaurule width#1\vss}}}
\def\phantomvrule#1{\vbox{\hbox to0pt{\vrule width\tableaurule height#1\hss}}}
\def\sqr{\vbox{%
  \phantomhrule\tableaustep
  \hbox{\phantomvrule\tableaustep\kern\tableaustep\phantomvrule\tableaustep}%
  \hbox{\vbox{\phantomhrule\tableauside}\kern-\tableaurule}}}
\def\squares#1{\hbox{\count0=#1\noindent\loop\sqr
  \advance\count0 by-1 \ifnum\count0>0\repeat}}
\def\tableau#1{\vcenter{\offinterlineskip
  \tableaustep=\tableauside\advance\tableaustep by-\tableaurule
  \kern\normallineskip\hbox
    {\kern\normallineskip\vbox
      {\gettableau#1 0 }%
     \kern\normallineskip\kern\tableaurule}%
  \kern\normallineskip\kern\tableaurule}}
\def\gettableau#1 {\ifnum#1=0\let\next=\null\else
  \squares{#1}\let\next=\gettableau\fi\next}

\title{Decoding the matrix: \\
Coincident membranes on the plane wave}

\author{%
Raphael Bousso and Aleksey L. Mints\\
Center for Theoretical Physics, Department of Physics\\
University of California, Berkeley, CA 94720-7300, U.S.A.\\
{\em and}\\
Lawrence Berkeley National Laboratory, Berkeley, CA 94720-8162, U.S.A.\\
E-mail: \email{bousso@lbl.gov, mints@socrates.berkeley.edu}}

\abstract{At the core of nonperturbative theories of quantum gravity
  lies the holographic encoding of bulk data in large matrices.  At
  present this mapping is poorly understood.  The plane wave matrix
  model provides a laboratory for isolating aspects of this problem in
  a controlled setting.

  At large boosts, configurations of concentric membranes become
  superselection sectors, whose exact spectra are known.  From the
  bulk point of view one expects product states of individual
  membranes to be contained within the full spectrum.  However, for
  non-BPS states this inclusion relation is obscured by Gauss law
  constraints.  Its validity rests on nontrivial relations in
  representation theory, which we identify and verify by explicit
  computation.}

\preprint{\hepth{0510121} \\ UCB-PTH-05/31 \\ LBNL-58946}

\begin{document}

\section{Introduction}
\label{sec-intro}

String theory has succeeded in producing nonperturbative formulations
of quantum gravity, such as Matrix theory~\cite{BFSS} and the AdS/CFT
correspondence~\cite{Mal97}.  Their application to many problems of
physical interest is encumbered by a number of restrictions and
technical challenges.  Among them is the requirement of an asymptotic
region where states are defined; such regions are not expected in a
realistic universe.  Another limitation is the lack of control over
non-supersymmetric excitations; again, this stands in the way of
treating problems such as cosmology, the interior of black holes, and
spacelike singularities.

However, great difficulties arise already in simpler situations,
because of the enormous complexity of the scrambled encoding of
physical data in a large matrix.  In this paper, we analyze an aspect
of the decoding problem, in the setting of the
Berenstein-Maldacena-Nastase (BMN) matrix model~\cite{BMN}.  

We work in a free limit where the only nontrivial aspect of the theory
lies in the matrix character of the degrees of freedom.  This will
allow us to focus very sharply on the kinematic problems arising from
the complexity of the matrices, without further complications from
dynamical evolution.  Moreover, it enables us to analyze and compare
full spectra, which are known exactly.  (As we shall show, the
encoding problem we pose simplifies significantly when restricted to
BPS states.)

The original Matrix theory~\cite{BFSS} was formulated for
asymptotically flat spacetime.  The understanding of its spectrum is
complicated by flat directions.  For example, the existence of
appropriate ground states has been difficult to demonstrate
rigorously, the distinction between single and multiple membranes is
problematic, and the description of transverse M5-branes has yet to be
understood (see Ref.~\cite{Tay01} for a review).

In the 11D plane wave~\cite{BMN} the curvature of the flux background
acts as a harmonic potential for transverse motion, lifting the flat
directions.  As a result, the spectrum can be computed exactly in a
certain perturbative limit~\cite{DSV1}.  In the same limit, the number
of distinct M2 branes is conserved and becomes a superselection
sector, despite the absence of a conserved charge.  Protected
states~\cite{DSV2} within the spectrum provide evidence~\cite{M5} for
the beautiful conjecture~\cite{BMN} that transverse M5 branes arise
nonperturbatively as the large $m$ limit of $m$ coincident M2 branes,
though the details of this transmutation remain mysterious.

In this paper we study the spectrum of a {\em finite\/} number $m$ of
coincident membranes in the plane wave matrix model.  We ask whether
and how a large matrix encodes the excitations of these $m$ physical
constituents.  Though we work in the perturbative limit where matrix
interactions are suppressed, this question is nontrivial due to gauge
constraints.  The matrix excitations are singlets under $U(m)$ and
correspond to products of traces.  Hence, individual operators in the
$m\times m$ matrices cannot be excited independently, as would be
required for the straightforward generation of the product spectrum.

We compare the spectrum obtained in the full matrix model with the
spectrum of a product of the constituent individual membranes.  The
full spectrum is constructed as follows. One begins with a set of
matrix creation operators, which act on the vacuum and form the same
representations as the one-particle states.  For a general state one
assigns up to $m$ matrices\footnote{This avoids overcounting due to
  trace relations; however, it does so only approximately.  Further
  subtleties arise because matrices inside the trace need not all be
  the same.  They are discussed in the main text and in
  Appendix~\ref{sec-rel}.}  to each trace, and then acts on the vacuum
with a product of an arbitrary number of traces.

On the other hand, the product spectrum of the constituent membranes
is constructed differently. Beginning with the same single-particle
spectrum, one assigns an arbitrary number of creation operators to
each membrane, and then takes the product of the $m$ single-membrane
spectra.

It is relatively straightforward to demonstrate that both spectra
contain the same set of BPS states.  But in the limit of a free
massive matrix model studied here, the two spectra can be compared in
far greater detail because all states are exact.

The symmetry group of this model is given by the supergroup $SU(4|2)$.
Each of the two spectra forms an infinite set of increasingly complex
irreducible representations of this group, and each spectrum is
completely specified by listing the representations it contains.  Thus
the assertion that the product spectrum is contained as a subset of
the full spectrum can be verified by comparing the multiplicities with
which each irreducible representation appears in the two spectra.

Let us outline and summarize the rest of this paper.  In
Sec.~\ref{sec-review}, we review the plane wave matrix model.  In
addition to the quartic interactions familiar from flat-space M(atrix)
theory, it has mass terms for all transverse directions (and
fermions).  It also contains a cubic interaction term that stabilizes
fuzzy sphere configurations via a Myers effect.

The vacuum of the matrix model is degenerate.  There is a $1/2$ BPS
ground state for each partition of $N$ into positive integers.  It
describes a set of concentric fuzzy spheres, which can be viewed as
rudimentary M2-branes.  The size of each sphere is proportional to the
corresponding integer in the partition.  The net M2 brane charge
vanishes for all configurations, and in general, the potential barrier
between vacua is finite.  Thus, excitations of the system can change
the number of fuzzy spheres.  In the large $N$ limit, this describes
processes such as two membranes merging into a single larger membrane.

At fixed $N$, the matrix model has an adjustable parameter, $\mu$,
corresponding to the steepness of the transverse potential.  In the
$\mu\to\infty$ limit the matrix model becomes free; only kinetic and
mass terms survive.  The barriers between vacua become infinite, so
each vacuum now forms the ground state of a superselection sector.  In
every sector, the Hamiltonian can be diagonalized explicitly, yielding
a collection of harmonic oscillators.  One obtains a discrete
spectrum~\cite{DSV1} by applying combinations of a finite set of
creation operators.

Thus, in the $\mu\to\infty$ limit, a given vacuum configuration of
fuzzy spheres persists for all excited states.  This makes it possible
(though not automatic) that the internal states of an individual fuzzy
sphere survive unmodified as additional fuzzy spheres are added to the
system.  If this is the case, then the spectrum should contain all
possible products of states associated with the individual
oscillations of each constituent.  For noncoincident spheres, this is
indeed guaranteed by the mutual independence of the operators
associated with the diagonal blocks representing each object inside
the full matrix.

Next, we turn to the question of whether the same holds true for $m$
coincident spheres.  In Sec.~\ref{sec-coin} we write down formal
expressions generating the representations contained in the product
spectrum, and, following Ref.~\cite{DSV2}, those in the full spectrum.
They are both given by sums of powers of the same one-particle
spectrum, but with different symmetrization patterns applied.  As a
consequence, the product representations cannot be straightforwardly
identified as a subset of the full spectrum, except in a few special
cases: if only one membrane is present (which is trivial), or in the
limit $m\to\infty$, which is problematic in the perturbative
regime.\footnote{For a toy matrix model with only one transverse
  direction, the two spectra would also be trivially related, because
  the symmetry group is $U(1)$ (see Sec.~\ref{sec-com}).  Hence, the
  problem cannot be simplified in this manner.}

Thus, for all finite $m>1$, the multiplicities of representations must
be worked out explicitly, and the results for the two spectra must be
compared.  Unfortunately, already the common single-particle spectrum
underlying both spectra can be arbitrarily rich and depends on the
size of the coincident fuzzy spheres.  In Sec.~\ref{sec-reduce} we
simplify our task by demonstrating that the product spectrum will be a
subset of the full spectrum, if the same is true for the much simpler
case where one pretends that the symmetry group is $U(\infty)$, and
the single particle spectrum is taken to consist only of the
fundamental of $U(\infty)$.\footnote{By $U(\infty)$ we mean $U(p)$
  with $p\to\infty$; the point is just to keep all Young diagrams
  regardless of the amount of antisymmetrization.}

However, even after this simplification one still faces the task of
comparing two infinite sets of irreducible representations for each
value of $m>1$.  In Sec.~\ref{sec-finite} we break up this task into
finite parts by noting that the total occupation number $F$ (the
number of operators applied to the vacuum) must be the same for two
representations to agree.  We find expressions generating only the
part of each spectrum corresponding to a given $F$.  Moreover, we show
that for each sector $F$, it suffices to check only a finite set of
values of $m$.  The $F$-sector of the full spectrum will contain the
representations found in the $F$-sector of the product spectrum if and
only if this holds for $2\leq m\leq F-3$.  We describe an algorithm
for carrying out this check, and compute the relevant representations
for all $F\leq 23$.  We find that the results are consistent with the
conjectured inclusion relation.  (A general proof is left to future
work.) We also prove that the number of BPS states matches precisely
for every $m$ and $F$.

In Sec.~\ref{sec-map}, we exhibit some of the nontrivial aspects of
the mapping of product states into the full spectrum.  We identify on
both sides all states without center-of-mass motion.  In the full
spectrum, this corresponds to the straightforward elimination of
certain reps~\cite{DSV2}.  For the product spectrum, however, it
requires taking linear combinations across different copies of a
representation.  This shows that in the natural bases, a given product
representation will generically be mapped into a nontrivial linear
combination of equivalent representations in the full spectrum.

In Sec.~\ref{sec-mlimit} we discuss subtleties that arise in the
M-theory limit, which corresponds to sending $N\to\infty$ while
holding $m$ and the longitudinal momentum fixed.  The question then
becomes whether M-theory, in the large $\mu$ limit, can describe $m$
noninteracting membranes of equal size.  

Various appendices contain additional information; in particular,
properties of plethysm are described in Appendix~\ref{sec-math}.

It is interesting that even the relatively simple decoding problem
studied here becomes extremely complex, in part due to Gauss law
constraints (which we treat exactly) and in part due to the subtleties
of trace relations (which we treat only approximately).  More powerful
techniques than we are aware of may allow a proof of the inclusion
relation conjectured in Sec.~\ref{sec-reduce}.  With a more
appropriate mathematical framework or by mapping to different physical
degrees of freedom, one might carry out the comparison of spectra
without approximations, and ideally, analytically.  Matrix encodings
should then become more transparent, and may well teach us lessons
about quantum gravity that go beyond the settings of asymptotically
flat and AdS spacetime.

\section{The plane wave: vacua, states, and representations}
\label{sec-review}

In this section we review the spectrum of the plane wave matrix model
in the $\mu\to\infty$ limit, and the representations it forms.  We
follow Refs.~\cite{DSV1,DSV2}, where more details can be found.

\subsection{Hamiltonian and vacua}

The $U(N)$ plane wave matrix model is given by the Hamiltonian
\begin{eqnarray} 
  H &=& R\ {\rm Tr}  \left( {1 \over 2} \Pi_A^2 - {1 \over 4} [X_A, X_B]^2 
    - {1 \over 2} \Psi^\top \gamma^A [X_A, \Psi] \right) \cr
  &+& {R \over 2} {\rm Tr}  \left( \left({\mu\over 
        3R}\right)^2  X_i^2 +  \left({\mu \over 6R}\right)^2 X_a^2 
  \right. \cr
  && \qquad \qquad \left. + 
  i {\mu \over 4R} \Psi^\top \gamma^{123} \Psi  + i {2\mu \over 3R} 
  \epsilon^{ijk} X_i X_j X_k \right)\ .   
\label{eq-model}
\end{eqnarray} 
Indices $A\ldots$ run from 1 to 9; $i\ldots$ run from 1 to 3; and
$a\ldots$ run from 4 to 9.  This model was proposed~\cite{BMN} to
describe M-theory on the maximally supersymmetric plane wave
background of eleven-dimensional supergravity,
\begin{eqnarray} 
  ds^2 &=& - 2 dx^+ dx^- + dx^A dx^A - \left( {\mu^2 
      \over 9} x^i x_i +  {\mu^2 \over 36} x^a x_a\right) (dx^+)^2~, \\
  F_{123+} &=& \mu~.
\label{eq-ppwave}
\end{eqnarray}
The M-theory limit is obtained by taking $N\to\infty$ while holding
$N/R$ fixed.  In this limit, all nonzero values of the parameter $\mu$
in Eq.~(\ref{eq-model}) must be physically equivalent since $\mu$ can
be set to unity by a coordinate redefinition in Eq.~(\ref{eq-ppwave}),
corresponding to a boost in the $x^\pm$ plane.

It was further proposed~\cite{Sus97} (originally in the context of the
flat space model~\cite{BFSS}) that the finite $N$ matrices describe
the discrete light cone quantization (DLCQ) of M-theory, in the sector
where the center of mass has $N$ units of longitudinal momentum.  This
corresponds to periodically identifying $x^- \cong x^-+2\pi R$ in
Eq.~(\ref{eq-ppwave}).  Then there exists a boost-invariant
dimensionless quantity, $\mu l_{\rm P}^2 /R$.\footnote{The Planck
  length is set to unity everywhere else in this paper.}  Thus, for
any finite $N$, the Hamiltonian (\ref{eq-model}) really describes a
one-parameter family of matrix models labelled by $\mu$.\footnote{We
  choose to keep $R$ fixed.  However, this is purely conventional.
  Arguably, it would be conceptually clearer not to introduce the
  coordinate artifact $\mu$ but to use only $R$ as a parameter.}

The classical supersymmetric solutions~\cite{BMN,DSV1} of the matrix
model are given by $X^a=0$ and
\begin{equation}
X^i = \frac{\mu}{3R} J^i~,
\end{equation}
where the $N\times N$ matrices $J^i$ form a representation (not
necessarily irreducible) of $SU(2)$:
\begin{equation}
[J^i,J^j] = i \epsilon^{ijk} J^k~.
\end{equation}
For any $l$ there is one irreducible $l\times l$ matrix representation
of $SU(2)$, so the vacua in question correspond to block-diagonal
solutions, one solution for each partition of $N$ into positive
integers.  

Physically, each block (of size $N_i$) represents a fuzzy sphere of
radius $r\sim \frac{\mu N_i}{6R}$.  Thus, a given partition $N=\sum
M_i N_i$ corresponds to a collection of concentric fuzzy spheres.  If
any of the multiplicities $M_i$ is greater than 1, the collection
includes coincident spheres.  For example, the partition $N=N$, the
``irreducible vacuum'', corresponds to a single membrane.  The
partition $N=1+1+\ldots+1$, the ``trivial vacuum'', contains $N$
coincident small fuzzy spheres, i.e., $N$ gravitons.

\subsection{Quantization and the free limit}
\label{sec-free}

The theory can be quantized by expanding about the classical vacua
\begin{equation}
X^i = \frac{\mu}{3R} J^i + Y^i~,
\end{equation}
and, for convenience, rescaling
\begin{equation}
Y^i \to \sqrt{\frac{R}{\mu}} Y^i,~
X^a \to \sqrt{\frac{R}{\mu}} X^a,~
t   \to \frac{t}{\mu}~.
\label{eq-rescale}
\end{equation}
We denote fermions by $\psi_{I\alpha}$, where $I$ is a fundamental
index of $SU(4)\sim SO(6)$, and $\alpha$ is a fundamental index of
$SU(2)\sim SO(3)$; see Ref.~\cite{DSV1} for the relation between these
complex eight-component spinors and the real 16 component spinor
$\Psi$.

The action for these fields consists of three separate, $\mu$- and
$R$-independent quadratic actions for $X^a$, $Y^i$, and
$\psi_{I\alpha}$, respectively; in addition there are cubic and
quartic interaction terms.  The interaction terms carry prefactors
given by positive powers of $R/\mu$.  Therefore, matrix interactions
are suppressed in the $\mu\to\infty$ limit.

The precise effective coupling depends on the vacuum~\cite{DSV1}.  We
will be interested mostly in partitions of the form $N=mn$, which
represent vacua consisting of $m$ coincident membranes, each with
$n$ units of longitudinal momentum.  For such vacua the effective
coupling behaves like an 't~Hooft coupling $\lambda=g^2 m$, with
$g^2=\left(\frac{R}{\mu n}\right)^3$.  But for any partition,
the effective coupling can be made arbitrarily small by taking $\mu$
sufficiently large.  Moreover, the height of the potential energy
barriers separating different vacua will scale with a positive power
of $\mu$.

In the limit $\mu\to\infty$ the theory thus becomes a free massive
matrix model.  Each vacuum becomes the ground state of a
superselection sector and can be quantized separately.  Explicit
diagonalization converts the Hamiltonian into a collection of free
harmonic oscillators.  They can be excited independently of one
another except where constrained by gauge invariance.  Physically,
they correspond to effects such as center of mass oscillations in the
transverse potential generated by the flux, and internal oscillations
of the fuzzy spheres about their spherical ground state.  We now turn
to a discussion of the resulting spectra.

\subsection{Spectra and representations}

The preceding discussion implies that the spectrum will consist of
towers of states of various harmonic oscillators.  Hence we can
associate to each state a {\em total occupation number\/}, $F$, given
by the total number of standard creation operators that are required
to create this state from the vacuum.  Note that $F$ is not in general
proportional to the energy of the state, since different oscillators
will have different energy spacing.

All states must fall into representations of the symmetry group of the
theory, the supergroup $SU(4|2)$.  The full set of physically
admissible representations contains continuous branches, as well as a
discrete subset of tensor representations.  The tensor representations
have integer highest weight and carry only fundamental indices of
$SU(4|2)$.  As is familiar from the representation theory of ordinary
unitary groups, each irreducible tensor representation is fully
characterized by the pattern in which its indices are symmetrized or
antisymmetrized, i.e., by a particular Young tableau whose number of
boxes agrees with the number of fundamental indices.  

Each such ``supertableau'' subsumes a set of $SU(4)\times SU(2)$
representations, as described in Refs.~\cite{BahBar82,DSV2}.  The
energy of an individual $SU(4)\times SU(2)$ representation is given
by\footnote{Due to the rescaling in Eq.~(\ref{eq-rescale}) energies
  are measured in units of $\mu$.}  $n_4/12+n_2/6$, where $n_4$ and
$n_2$ are the number of boxes assigned to the $SU(4)$ and $SU(2)$
indices respectively~\cite{DSV2}.  Note that because the relevant
supersymmetries do not commute with the Hamiltonian, different bosonic
representations contained in the same superrepresentation will have
different energy.  But generally the energy of states associated with
a given superrepresentation is of order the number of boxes.

The sum of the irreducible $SU(4|2)$ representations that are actually
present in the spectrum of the theory, for a given partition $P$, will
be denoted $R(P)$.  $P$ is specified as a list of nonincreasing
integers or a Young diagram.  We use the notation $P=(N_1^{M_1}\cdots
N_L^{M_L})$ with $M_i$, $N_i$ positive integers and
$N_1>N_2>\ldots>N_L$.  This describes a Young diagram with $M_i$ rows
of length $N_i$, corresponding to a partition of $N=\sum M_iN_i$ into
integers given by the length of each row.  (See
Appendix~\ref{sec-math} for further details and notation.)  Thus,
$R(N)$ refers to the representations of the irreducible vacuum,
$R(4~3^2)$ to the partition of $N=10$ into $4+3+3$, and $R(n^m)$ to
the partition of $N=mn$ into $n+n+\ldots+n$.

It was shown in Ref.~\cite{DSV2} that at $\mu\to\infty$, for any
partition $P$, the plane wave matrix model contains only tensor
representations.  This implies that the set $R(P)$ can be specified by
an (infinite) list of Young tableaux.  Moreover, it can be shown that
the generators of $SU(4|2)$ leave the total excitation number, $F$,
invariant.  This means that we can break up $R(P)$ into finite chunks
according to occupation number.  We will denote the (finite) set of
representations formed by the states with occupation number $F$ by
$R(P)_F$.

After diagonalizing the Hamiltonian at $\mu\to\infty$, one can
associate a creation operator to each harmonic oscillator.  The full
spectrum arises by acting on the vacuum with combinations of these
creation operators.  Since creation operators commute,\footnote{For
  operators in representations of a supergroup, this statement
  subsumes the commutativity of operators in bosonic subgroups and the
  anticommutativity of the fermionic operators.}  this suggests that
the full spectrum will consist of symmetrized products of
representations in $R(P)_1$, the single particle sector, and that
$R(P)$ can be obtained by expanding all such tensor products using the
Littlewood-Richardson rule.  However, the matrix model is gauged, so
physical states must be invariant under $U(N)$ gauge transformations.
This Gauss law constraint takes the form
\begin{equation}
i[X^i,\Pi^i] + i[X^a,\Pi^a] + 2\psi_{I\alpha}\psi^{\dagger I\alpha} = 0~.
\label{eq-gauss}
\end{equation}
The details of its implementation depends on the vacuum $P$, as we
will discuss next.

\subsection{Single membrane}

First, consider the states built on the irreducible vacuum, the
partition with a single element $N$.  In this case the specification
of the solution completely fixes the gauge.  The $N^2-1$ nontrivial
generators of $U(N)$ gauge transformations show up as a set of
zero-mode oscillators when the Hamiltonian is diagonalized.  These
gauge orbit directions are not associated with creation operators.
The states obtained from creation operators associated with the
remaining sets of oscillators are then automatically physical and
gauge-fixed.

The physical creation operators carry an even number of
antisymmetrized $SU(4|2)$ indices\footnote{Except for Sec.~\ref{sec-com}
  we do not use annihilation operators.  Hence we drop the $\dagger$
  symbol on creation operators in the interest of a less cluttered
  notation.}
\begin{equation}
  a_{[I_1I_2]},~a_{[I_1I_2I_3I_4]},~\ldots,~a_{[I_1\cdots I_{2N}]}~.
\label{eq-ns}
\end{equation}
Hence the single particle states ($F=1$) form single-column
representations of $SU(4|2)$ with $2,~4,~\ldots,~2N$ boxes:
\begin{equation}
R(N)_1 ={\tiny \yngSLASH(1,1) \, \oplus \, \yngSLASH(1,1,1,1)
\, \oplus \, \ldots \, \oplus \,
\left. \yngSLASH(1,1,1,1,1,1) \right\}2N}~.
\label{eq-nsrep}
\end{equation}

The full spectrum is generated by all combinations of physical
creation operators.  Since the gauge is already completely fixed,
there are no restrictions on such combinations.  Hence, the full list
of irreps, $R(N)$, is given by
\begin{equation}
  R(N) = \sum_F R(N)_F = \sum_F \mathsf{sym}~ R(N)_1^F~.
\label{eq-na}
\end{equation}
This can also be written in the form 
\begin{equation}
  R(N) = \bigoplus_{\{n_i\}} \left[
    {\tiny 
      \mathsf{sym}~ \left(\,\yngSLASH(1,1) \, \right)^{n_1} \, \otimes \,
      \mathsf{sym}~ \left(\, \yngSLASH(1,1,1,1) \, \right)^{n_2} \, 
      \otimes \,      \ldots \, \otimes \,
      \mathsf{sym}~ \left(\, 
        \left. \yngSLASH(1,1,1,1,1,1) \right\}2N \, \right)^{n_{N}}    }
  \right]~.
\end{equation}
Symmetrization is necessary for powers of identical representations in
order to avoid overcounting states, since exchange of the complete set
of indices between any two operators in the power leaves the state
invariant.

\subsection{Noncoincident membranes}
\label{sec-nonc}

A partition of $N$ into $L$ {\em different\/} integers $N_i$
corresponds to $L$ noncoincident concentric spheres of radii
proportional to $N_i$.  This configuration preserves a portion of the
gauge group $U(N)$, namely $U(1)^L$ (including the trivial $U(1)$
associated with the center of mass), under which physical states must
be invariant.

In this configuration the matrices break up into $L^2$ $(N_i \times N_j)$
blocks, which
are separately diagonalized.  The $i$-th diagonal block carries the
same set of oscillators as the single membrane vacuum with $N=N_i$:
\begin{equation}
  a^{ii}_{[I_1I_2]},~a^{ii}_{[I_1I_2I_3I_4]},
  \ldots,~a^{ii}_{[I_1\cdots I_{2N_i}]}~;~~i=1,\ldots,L~.
\label{eq-single}
\end{equation}
Again, the $N_i^2-1$ zero modes correspond to gauge orbits and are
discarded.  The remaining operators in the diagonal blocks do not
break the $U(1)^L$ invariance of the vacuum and hence can be applied
without restriction.  The resulting spectrum forms precisely the
representations in the tensor product $\prod R(N_i)$ of the
individual membranes' sets of representations.  

Thus, the noncoincident multimembrane vacuum trivially contains the
degrees of freedom of its constituents:
\begin{equation}
  R(N_1\cdots N_L)  \supseteq \prod_{i=1}^L R(N_i)~,
\label{eq-noncsup}
\end{equation}
where $R(N_i)$ is given by Eq.~(\ref{eq-na}).  Indeed, this
relation holds at each individual level of the total occupation
number, as it should:
\begin{equation}
  R(N_1\cdots N_L)_F  \supseteq \left[\prod_{i=1}^L R(N_i)\right]_F~,
\label{eq-noncsupf}
\end{equation}
where we recall that the final index denotes the restriction of the
full spectrum to representations with occupation number $F$.

The full spectrum contains additional representations coming from the
off-diagonal blocks, each of which has its own set of oscillators.
The associated operators again carry purely antisymmetrized $SU(4|2)$
indices.  The number of indices (and thus the number of boxes in the
corresponding single-column representation) ranges in steps of 2 from
$|N_i-N_j|+2$ to $N_i+N_j$:
\begin{equation}
  a^{ij}_{[I_1\cdots I_{|N_i-N_j|+2}]},~a^{ij}_{[I_1\cdots I_{|N_i-N_j|+4}]},~
  \ldots,~a^{ij}_{[I_1\cdots I_{N_i+N_j}]}~;~~i\neq j~.
\label{eq-nonc}
\end{equation}
Since the off-diagonal blocks transform in the bifundamental of
$U(1)_i\times U(1)_j$, their operators must be combined to form
$U(1)^L$ invariants (e.g., $a^{12}a^{23}a^{31}|0\rangle$).  

These off-diagonal excitations connect pairs of membranes by tubes
wrapping the eleventh dimension~\cite{AhaSon96}.  In the
decompactification limit ($N\to\infty$, keeping $N_i/N$ fixed), such
tubes become infinitely massive.  Indeed, the energy of the lightest
off-diagonal state is of order $|N_i-N_j|$ and diverges in this limit.

\subsection{General vacuum}

A general partition $(N_1^{M_1}\cdots N_L^{M_L})$ corresponds to $L$
concentric stacks of coincident membranes.  The $i$-th stack contains
$M_i$ membranes of size $N_i$ sitting on top of one another.  The
general vacuum state is invariant under $U(M_1)\times\ldots\times
U(M_L)$.  For physical excitations the same invariance is enforced by
the Gauss law constraint.  In general, this constraint significantly
complicates the explicit construction of the spectrum and especially
its comparison with the constituent spectra.  We now turn to a
detailed analysis of this problem.

We conclude this review section by pointing once more to
Refs.~\cite{BMN,DSV1,DSV2}, where more detailed descriptions and
derivations can be found.

\section{Coincident membranes}
\label{sec-coin}

In this section we present the spectra to be compared.  We begin by
reducing the most general problem to a simpler question.

We would like to know whether the spectrum of a general membrane
configuration $(N_1^{M_1}\cdots N_L^{M_L})$ contains the product
states arising from its constituents, i.e., whether a relation
analogous to Eq.~(\ref{eq-noncsupf}) holds in the general case.  By
arguments completely analogous to those given in Sec.~\ref{sec-nonc}
for noncoincident membranes, it is trivial to show that
\begin{equation}
  R(N_1^{M_1}\cdots N_L^{M_L}) \supseteq \prod_{i=1}^L  R(N_i^{M_i})~,
\end{equation}
where $R(N_i^{M_i})$ refers to the spectrum that the $i$-th stack of
coincident branes would have by itself.  Thus, the independence of the
degrees of freedom of one stack of membranes from those of another is
manifest.

Then the remaining problem is to identify the individual membranes
within a single stack.  This requires the comparison of two spectra.
The first consists of the symmetrized product states of $m$ identical
fuzzy spheres of size $n$.  The second is the spectrum $R(n^m)$ built
on the matrix theory vacuum corresponding to the partition of $N=mn$
into $m$ coincident fuzzy spheres.

\subsection{Product spectrum}

The product spectrum is the $m$-th tensor power of the single membrane
spectrum, restricted to representations compatible with the bosonic
exchange symmetry of identical membranes:
\begin{equation}
  R^{\rm prod}(n^m) =  \mathsf{sym}~ R(n)^m = 
  \mathsf{sym}~ \left[\sum_{g=0}^\infty \mathsf{sym}~ R(n)_1^g\right]^m~,
\label{eq-prodsp1}
\end{equation}
where Eq.~(\ref{eq-na}) has been used for the second equality.

Using properties of $\mathsf{sym}$ described in
Appendix~\ref{sec-math}, one can eliminate the trivial representation
$g=0$ in exchange for introducing another summation:
\begin{equation}
  R^{\rm prod}(n^m) = \sum_{f=0}^m
  \mathsf{sym}~ \left[\sum_{g=1}^\infty \mathsf{sym}~ R(n)_1^g\right]^f~.
\label{eq-prodsp2}
\end{equation}

\subsection{Full spectrum}
\label{sec-full}

We turn to the spectrum of the full matrix model in the partition
$(n^m)$.  In this vacuum, the matrices break up into $m^2$ square
blocks, each of which gauge-fixes $n^2-1$ of the $U(N)$ generators.
The vacuum is invariant under the remaining $U(m)$ gauge group, with
$m^2-1$ nontrivial generators.  The Gauss law constraint,
Eq.~(\ref{eq-gauss}), demands that physical states are likewise
invariant under $U(m)$.

In each block the Hamiltonian can be diagonalized~\cite{DSV1}, leading
to the set of operators found for the single membrane vacuum with
$N=n$.  This includes $n^2-1$ zero modes corresponding to gauge
orbits, which are tossed out.  Combining all blocks, one can think of
each physical operator as a $m\times m$ matrix operator, e.g.,
\begin{equation}
A_{[I_1I_2]} = 
\left(
\begin{matrix}
a^{11}_{[I_1I_2]} & a^{12}_{[I_1I_2]} & \ldots & a^{1m}_{[I_1I_2]} \\
a^{21}_{[I_1I_2]} & a^{22}_{[I_1I_2]} & \ldots & a^{2m}_{[I_1I_2]} \\
\vdots & \vdots & \ddots & \vdots \\
a^{m1}_{[I_1I_2]} & a^{m2}_{[I_1I_2]} & \ldots & a^{mm}_{[I_1I_2]}
\end{matrix}
\right)~.
\end{equation}
$U(m)$ invariant objects arise from traces of products of these
matrices.  Hence, physical states are obtained by acting on the vacuum
with products of traces of products of matrix creation operators,
e.g.:
\begin{equation} 
  {\rm Tr} (A_{[I_1I_2]}A_{[J_1\cdots J_6]}) 
  {\rm Tr} (A_{[K_1\cdots K_8]}) |0\rangle~.
\end{equation}

Next we discuss the $SU(4|2)$ representations formed by the physical
states~\cite{DSV2}.  Each matrix $A$ is a single column representation
since the indices are totally antisymmetrized.  This is unaffected by
taking the trace, so the single particle states, ${\rm Tr}~A
|0\rangle$, form the same representations as those of a single
membrane with $N=n$:
\begin{equation}
R(n^m)_1 = R(n)_1~.
\label{eq-r1}
\end{equation}
(The supertableaux in $R(n)_1$ are shown in Eq.~\ref{eq-nsrep}.)

The full spectrum will be given by tensor products of these
representations, subject to appropriate symmetrization.  Matrices
within a single trace do not commute, so total symmetrization would be
too restrictive.  Instead, the allowed representations must follow a
symmetrization pattern, denoted $\mathsf{cyc}$, that is compatible
with the cyclic symmetry of the trace.  Its properties are described
in greater detail in Appendix~\ref{sec-math}, where it is shown that
$\mathsf{cyc}$ corresponds to the sum of all plethysms by Young
diagrams with an odd number of rows.\footnote{We thank M.~Van
  Raamsdonk for pointing this out to us.}  In particular, the
representations generated by $\mathsf{cyc}$ always contain those
generated by $\mathsf{sym}$ as a subset.

This suggests that single trace states with $g$ operators will form
representations $\mathsf{cyc~} R(n)_1^g$.  However, there is a danger
of overcounting due to trace relations.  Traces containing more than
$m$ equal factors are equal to products of smaller traces.
Correspondingly we shall require $g\leq m$, leading to single-trace
representations
\begin{equation}
  R'_{\rm single-trace}(n^m) =
  \sum_{g=1}^m  \mathsf{cyc~} R(n)_1^g~.
\label{eq-singletrace}
\end{equation} 

A prime has been added to indicate that $R'_{\rm single-trace}$ is a
truncated spectrum.  This is because our restriction $g\leq m$ is
actually somewhat excessive.  There are many different irreducible
representations, and each contains several different matrix operators.
The standard trace relations apply rigorously only to traces of a
power of a single operator.  Indeed, independent states with more than
$m$ operators inside the trace can exist if not all the operators are
equal.  However, it is cumbersome\footnote{See the rather intricate
  analysis for even the simple case $n=1$, $m=2$, $F=3$ in
  Ref.~\cite{DSV2}.  The single trace spectrum in this case is given
  by the plethysm $R(1)^{\otimes(1^3)}$ [which could be denoted
  loosely as ($\mathsf{cyc~}-\mathsf{sym~}) R(1)^3$].  Moreover, a
  general state is not allowed to contain more than one such trace.
  (There is a minor misleading statement in Ref.~\cite{DSV2}
  suggesting that the full $\mathsf{cyc~}$ is present; note also that
  the $U(1)$ degrees of freedom are suppressed there.)} to extract the
allowed combinations.  Hence, though we only wish to show that the
product of individual membrane spectra is contained in the full
spectrum $R(n^m)$, we will in fact argue that the stronger statement
is true that it is contained already among the representations
computed from the truncated single-trace spectrum $R'_{\rm
  single-trace}$.

A general state contains multiple traces.  These factors commute and
must be symmetrized.  The full spectrum (up to the truncation
introduced earlier) is thus
\begin{eqnarray} 
  R'(n^m) &=& \sum_{f=0}^\infty \mathsf{sym}~ 
  \left[R'_{\rm single-trace}(n^m)\right]^f\\
  &=& \sum_{f=0}^\infty \mathsf{sym}~ 
  \left[\sum_{g=1}^m  \mathsf{cyc~} R(n)_1^g\right]^f~,
\label{eq-fullsp}
\end{eqnarray} 
{\em assuming that no trace relations arise for multitrace states}. 
As explained in Appendix~\ref{sec-rel}, one expects this assumption to
be well satisfied for occupation numbers $F\ll$ dim$R(n)_1 m^2$.  Note
that this upper bound grows rapidly with $n$ and diverges in the
M-theory limit.  Nevertheless, it is important to keep in mind that
our expression for $R'$ should be expected to contain a small number
of redundant representations.  It is possible that eliminating these
states could spoil the inclusion relation we find in
Sec.~\ref{sec-finite}.  In light of the evidence presented in
Appendix~\ref{sec-rel} we consider this possibility remote.

\subsection{Assessment}
\label{sec-analysis}

We would like to determine whether the representations
formed by the product spectrum of $m$ membranes,
Eq.~(\ref{eq-prodsp2}), is a subset of the representations formed by
the full spectrum of the $U(m)$ theory, Eq.~(\ref{eq-fullsp}), as one
would expect physically; i.e., is
\begin{equation}
R'(n^m)\supseteq R^{\rm prod}(n^m)~?
\label{eq-qu}
\end{equation}

The two expressions given above differ in two ways.  The first is
quite trivial: as explained in Appendix~\ref{sec-math}, a
cyclically symmetrized power (which arises for $U(m)$) always contains
the corresponding totally symmetrized power (arising in the product
spectrum) as a subset.  One can thus consider truncating the $U(m)$
spectrum further by replacing the $\mathsf{cyc}$ with $\mathsf{sym}$
in Eq.~(\ref{eq-fullsp}).\footnote{This modification
  includes as a special consequence the earlier truncation $R\to R'$
  introduced in Eq.~(\ref{eq-singletrace}), since the usual trace
  relations do apply to traces of totally symmetrized products.}
This yields the spectrum
\begin{equation}
  R''(n^m) =
  \sum_{f=0}^\infty \mathsf{sym}~ 
  \left[\sum_{g=1}^m  \mathsf{sym~} R(n)_1^g\right]^f~.
\label{eq-fullsps}
\end{equation} 

The spectra obey $R\supseteq R'\supseteq R''$.  It would be nice to
continue this sequence of inclusions to $R^{\rm prod}$.  However, even
after replacing cyclic with total symmetrization, a crucial difference
between $R''$ and $R^{\rm prod}$ remains: the exchange of the upper
limits, $m$ and $\infty$, of the double sum.

Clearly the spectra $R''$ and $R^{\rm prod}$ agree in the limit of
infinitely many membranes, $m\to\infty$.\footnote{In order to take
  this limit while keeping the coupling $\lambda$ small and fixed,
  $\mu$ must be scaled like $m^{1/3}$ (see Sec.~\ref{sec-free}).}
Also, one easily verifies that they reduce to the same expression for
the case of a single membrane, $m=1$.  But for all other values of $m$
the spectra $R''$ and $R^{\rm prod}$ differ, and it is not apparent
whether the former contains the latter.  Hence a more detailed
analysis is needed, which we begin in the next section.

\section{Irrelevance of the single particle spectrum}
\label{sec-reduce}

It may appear that the validity of Eq.~(\ref{eq-qu}) depends on the
detailed structure of $R(n)_1$, the single-particle spectrum.  It
contains various irreducible tensor representations of the supergroup
$SU(4|2)$ (see Eq.~(\ref{eq-nsrep})).  In particular, in the M-theory
($n\to\infty$) limit, $R(n)_1$ contains an arbitrarily large number of
different irreducible terms.  This would appear to complicate the
comparison of the spectra $R'$ and $R^{\rm prod}$, which are
constructed from symmetrized powers of $R(n)_1$.  Moreover, it might
seem that the validity of Eq.~(\ref{eq-qu}) could depend on the value
of $n$.

However, in this section we will show the problem simplifies
considerably: it suffices to show that $R'\supseteq R^{\rm prod}$ for
the case where $R(n)_1$ is assumed to contain only a single,
fundamental representation of the unitary group, which we shall think
of as having infinite degree, $U(\infty)$.  (Note that this does not
even correspond to any choice of parameters for the matrix model at
hand.)  In particular, we are guaranteed that Eq.~(\ref{eq-qu}) holds
for all $n$ and in the M-theory limit, if this simpler case can be
proven.

This simplification arises as follows.  Beginning with $R'$,
write the symmetric and cyclic powers in Eq.~(\ref{eq-fullsp}) as
plethysms (see Appendix~\ref{sec-math}):
\begin{equation}
  R'(n^m) = \sum_{f=0}^\infty\left[\sum_{g=1}^m
    R(n)_1^{\otimes ({\rm oddrow}_g)}\right]^{\otimes(f)}~,
\end{equation}
where the partition $(f)$ denotes the Young diagram corresponding to
complete symmetrization (one row of length $f$).  As explained in
Appendix~\ref{sec-math}, $({\rm oddrow}_g)$, which denotes the sum of
all Young diagrams with $g$ boxes and an odd number of rows, is the
plethysm corresponding to a cyclically symmetrized power.  

Since plethysm is distributive in the ``exponent'', we have
\begin{equation}
  R'(n^m) = \left[R(n)_1^{\otimes\sum_{g=1}^m
      ({\rm oddrow}_g)}\right]^{\otimes\sum_{f=0}^\infty (f)}~.
\end{equation}
Associativity of plethysm implies
\begin{equation}
  R'(n^m) = R(n)_1^{\otimes\left[\left(\sum_{g=1}^m ({\rm oddrow}_g)\right)
      ^{\otimes\sum_{f=0}^\infty (f)}\right]}~.
\end{equation}
Similarly, $R^{\rm prod}$ can be written in the form
\begin{equation}
  R^{\rm prod}(n^m) = R(n)_1^{\otimes\left[\left(\sum_{g=1}^\infty (g)\right)
      ^{\otimes\sum_{f=0}^m (f)}\right]}~.
\end{equation}
Hence Eq.~(\ref{eq-qu}) will hold, irrespective of $R(n)_1$, if the
corresponding relation holds for the $U(\infty)$ representations
contained in the exponents.

Thus, the problem reduces to the question of whether
\begin{equation}
S'(m)\supseteq S^{\rm prod}(m)~,
\end{equation}
where
\begin{eqnarray}
S'(m) & \equiv & \left(\sum_{g=1}^m (\mbox{oddrow}_g)\right)
  ^{\otimes\sum_{f=0}^\infty (f)}~,
\label{eq-sprime}\\
S^{\rm prod}(m) & \equiv & \left(\sum_{g=1}^\infty(g)\right)
  ^{\otimes\sum_{f=0}^m (f)}~.
\label{eq-sprod}
\end{eqnarray} 
Unlike $R'$ and $R^{\rm prod}$, these expressions are independent of
$n$; the argument $m$ here is not a partition but denotes the number
of (arbitrary-size) coincident membranes in the stack.

The $S'$ and $S^{\rm prod}$ coincide with $R'$ and $R^{\rm prod}$
precisely if we set $R(n)_1=\tableau{1}$, as advertised above.  This
is because $\mathsf{sym}~\tableau{1}^g$ gives precisely the one-row
Young diagram of length $g$, here denoted $(g)$; and
$\mathsf{cyc}~\tableau{1}^g$ gives the odd-row Young diagrams with $g$
boxes.  In this sense we have reduced the problem to the simple case
where the single particle states lie only in the fundamental of
$U(\infty)$.

A similar simplification can be derived involving $R''$, the spectrum
obtained from the truncation $\mathsf{cyc} \to \mathsf{sym}$.  Writing
$R''$ as a plethysm of $R(n)_1$, one obtains $R''(n^m)=R(n)_1^{\otimes
  S''(m)}$, where
\begin{equation}
S''(m) \equiv \left[\sum_{g=1}^m (g)\right]
  ^{\otimes\sum_{f=0}^\infty (f)}~.
\end{equation}
Clearly, Eq.~(\ref{eq-qu}) follows if it can be shown that
$R''(n^m)\supseteq R^{\rm prod}(n^m)$, which in turn is implied by
\begin{equation}
S''(m)\supseteq S^{\rm prod}(m)~. 
\label{eq-q}
\end{equation}
We will provide some evidence below that this relation does hold.

\section{Comparison at finite occupation number}
\label{sec-finite}

In the absence of a proof of Eq.~(\ref{eq-q}), we would like to assess
its plausibility by verifying parts of the relation explicitly.  This
can be done systematically by restricting both $S''(m)$ and $S^{\rm
  prod}(m)$ to states with a given occupation number $F$.  The
occupation number of an irreducible $U(\infty)$ representation in
$S''$ or $S^{\rm prod}$ is the number of boxes in the corresponding
$U(\infty)$ Young diagram.\footnote{Acting as a plethysm, a
  representation with $U(\infty)$ occupation number $F$ produces a
  subset of the $F$-th power of the single particle states $R(n)_1$.
  Hence, the resulting occupation number in the $R$ spectra will again
  be $F$.  This will become quite explicit in the examples below.}
Clearly, the inclusion relation Eq.~(\ref{eq-q}) can only be true if
it holds separately for each $F$:
\begin{equation}
S''(m)_F\supseteq S^{\rm prod}(m)_F~. 
\label{eq-qf}
\end{equation}

\subsection{Auxiliary Young diagrams}
\label{sec-auxiliary}

Let us expand
\begin{eqnarray}
S''(m)_F & = & \bigoplus_{P(F)} c''(m)_{P(F)} P(F)~, \cr
S^{\rm prod}(m)_F & = & \bigoplus_{P(F)} c^{\rm prod}(m)_{P(F)} P(F)~,
\label{eq-expand}
\end{eqnarray} 
where the sum is over all partitions of $F$, i.e., all Young diagrams
with $F$ boxes.  Thus, at finite $F$ we would like to verify that the
multiplicities of each representation obey
\begin{equation}
c''(m)_{P(F)} \geq c^{\rm prod}(m)_{P(F)}
\label{eq-coeff}
\end{equation}
for all $P(F)$.  This requires a systematic way of isolating the terms
in $S''$ and $S^{\rm prod} $ that lead to $F$-box representations.
Then the multiplicities can be tallied up and compared.

Consider the example $m=3$.  One can write the expressions for $S$ in
the form
\begin{eqnarray}
  S^{\rm prod}(3) & = & ~~~~~~~\bullet \cr
  & \oplus & ~~~~~~\left(\tableau{1}\oplus\tableau{2}
    \oplus\tableau{3}\oplus\tableau{4}\oplus\cdots\right) \cr
  & \oplus & \mathsf{sym}~ \left(\tableau{1}\oplus\tableau{2}
    \oplus\tableau{3}\oplus\tableau{4}\oplus\cdots\right)^2 \cr
  & \oplus & \mathsf{sym}~ \left(\tableau{1}\oplus\tableau{2}
    \oplus\tableau{3}\oplus\tableau{4}\oplus\cdots\right)^3~,
\label{eq-m3}
\end{eqnarray}
where $\bullet$ stands for the trivial representation, i.e., the
vacuum.  The first and second row need no evaluation.  For the higher
powers, the $\mathsf{sym}$ ensures that each (commuting) product of
representations occurs with multiplicity 1.  For example the third row
evaluates to
\begin{equation}
\mathsf{sym}~\tableau{1}^2 \oplus (\tableau{2}\cdot\tableau{1}) \oplus 
\mathsf{sym}~\tableau{2}^2 \oplus (\tableau{3}\cdot\tableau{1}) \oplus 
(\tableau{3}\cdot\tableau{2}) \oplus \mathsf{sym}~\tableau{3}^2 \oplus 
\cdots
\label{eq-sq}
\end{equation}
Imagine writing the factors in each term underneath one another.  This
yields a set of ``auxiliary Young diagrams''.  They do not themselves
correspond to representations in the expansion of Eq.~(\ref{eq-m3}),
but they help us keep track of the terms that will appear there.  For
example the terms in Eq.~(\ref{eq-sq}) are in one-to-one
correspondence with auxiliary diagrams
\begin{equation}
\tableau{1 1}~,~\tableau{2 1}~,~\tableau{2 2}~,~\tableau{3
  1}~,~\tableau{3 2}~,~\tableau{3 3}~,~\cdots
\end{equation}
Note that every Young diagram with 2 rows occurs precisely once.
Similarly, the 3-factor product terms arising from the last row of
Eq.~(\ref{eq-m3}) are in correspondence with all 3-row auxiliary Young
diagrams.

Let us summarize and generalize the above.  Expanding the powers of
the infinite sum in $S^{\rm prod}(m)$ into a sum over products, one
finds that the resulting terms are in one-to-one correspondence with
all auxiliary Young diagrams with up to $m$ rows.

For example the auxiliary diagram $\tableau{4 2 2}$ (which will be
considered only if $m\geq 3$) corresponds to the term $\tableau{4}
\cdot \mathsf{sym}~ \tableau{2}^2$.  Further evaluation of this
product yields a sum of various irreducible $U(\infty)$
representations, each with $F=4+2+2=8$ boxes.

In order to obtain the multiplicities for all $F$-box representations
in $S^{\rm prod}(m)$, one can thus proceed as follows.  Write down all
$F$-box auxiliary Young diagrams with up to $m$ rows; for each,
evaluate the symmetric product of the rows; sum up the results.  In
other words:
\begin{equation}
  S^{\rm prod}(m)_F= \bigoplus_{P(F/m) = (q_1^{p_1}\cdots q_l^{p_l})}~ 
  \prod_{i=1}^l \mathsf{sym}~(q_i)^{p_i}~,
\label{eq-sprodmf}
\end{equation}
where the sum is over all partitions $P(F/m)$ of $F$ with up to $m$
terms (i.e., $\sum_{i=1}^l p_i\leq m$).

For $S''(m)$, an analogous argument leads to the following
prescription: Write down all $F$-box auxiliary Young diagrams with up
to $m$ {\em columns}; for each, evaluate the symmetric product of the
rows; sum up the results:
\begin{equation}
  S''(m)_F= \bigoplus_{P(F\backslash m) = (q_1^{p_1}\cdots q_l^{p_l})}~ 
  \prod_{i=1}^l \mathsf{sym}~(q_i)^{p_i}~,
\label{eq-sfullmf}
\end{equation}
where the sum is over all partitions $P(F\backslash m)$ of $F$ with no
term greater than $m$ (i.e., $q_1\leq m$).  This will produce all
$F$-box representations and their multiplicities $c''(m)_{P(F)}$.

\subsection{Interpretation}
\label{sec-interpretation}

The above expressions, Eqs.~(\ref{eq-sprodmf}) and (\ref{eq-sfullmf})
have a straightforward combinatorial interpretation, even in the
original context of the $R$-spectra.  

The expression for $S^{\rm prod}(m)_F$ corresponds to the problem of
distributing $F$ creation operators, each in the fundamental of
$U(\infty)$, among $m$ identical objects.  The distribution is
described first by the partitions of $F$ with no more than $m$ terms
($\oplus_{P(F/m)}$).  For each partition, there are many different
states corresponding to a choice of specific operators from the
fundamental multiplet.  However, the resulting representation is not
simply $\tableau{1}^{F}$.  The $q_i$ operators assigned to one
particle commute and must be totally symmetrized.  Thus the resulting
states lie in the representation $\mathsf{sym}~\tableau{1}^{q_i}=
(q_i)$ (the one-row diagram with $q_i$) boxes.  The representations
for the different particles must be tensored together
($\prod_{i=1}^l$), taking care to symmetrize between particles that
are not distinguished by different oscillator number
($\mathsf{sym}~(q_i)^{p_i}$).

The expression for $S''(m)_F$ can be similarly understood.  In fact,
consider the more general case of
\begin{equation} 
  S'(m)_F= \bigoplus_{P(F\backslash m) = (q_1^{p_1}\cdots q_l^{p_l})}~
  \prod_{i=1}^l \mathsf{sym}~({\rm oddrow}_{q_i})^{p_i}~,
\label{eq-sprimemf}
\end{equation}
which arises analogous to Eq.~(\ref{eq-sfullmf}) by restricting
Eq.~(\ref{eq-sprime}) to the sector with occupation number $F$.  This
describes the distribution of $F$ matrix operators among an arbitrary
number of traces, assigning no more than $m$ operators to each trace
($\oplus_{P(F\backslash m)}$).  The operators assigned to one trace
are cyclically symmetric.  The resulting states lie in
$\mathsf{cyc}~\tableau{1}^{q_i} = ({\rm oddrow}_{q_i})$.  Different
traces commute and must be symmetrized.  ($S''(m)_F$ is the truncation
of this spectrum to those states which are totally symmetrized within
each trace.)

With this interpretation, it is clear that these expressions can be
lifted back up to the original case where the single-particle
representation is more complicated.  One replaces the fundamental of
$U(\infty)$ with the actual single-particle representation of the
symmetry group:
\begin{eqnarray}
  R'(n^m)_F &=& \bigoplus_{P(F\backslash m) = (q_1^{p_1}\cdots q_l^{p_l})}~
  \prod_{i=1}^l \mathsf{sym}~(\mathsf{cyc}~R(n)_1^{q_i})^{p_i}~;
\label{eq-ijk} \\
  R''(n^m)_F &=& \bigoplus_{P(F\backslash m) = (q_1^{p_1}\cdots q_l^{p_l})}~
  \prod_{i=1}^l \mathsf{sym}~(\mathsf{sym}~R(n)_1^{q_i})^{p_i} ~;
\label{eq-jkl}\\
  R^{\rm prod}(n^m)_F &=& \bigoplus_{P(F/m) = (q_1^{p_1}\cdots q_l^{p_l})}~ 
  \prod_{i=1}^l \mathsf{sym}~(\mathsf{sym}~R(n)_1^{q_i})^{p_i}~.
\label{eq-klm}
\end{eqnarray}

\subsection{BPS states}
\label{sec-bps}

There is an important subset of states which are protected as $\mu$ is
decreased away from the $\mu\to\infty$ limit.  These ``doubly
atypical'' states are analogues of BPS states~\cite{DSV2}, and they
come from superrepresentations with exactly two equal rows of
arbitrary length:
\begin{equation}
{\tiny \yngSLASH(7,7)}~.
\end{equation}
Such representations can only be generated from plethysms of $\,{\tiny
  \yngSLASH(1,1)}\,$; no other diagrams in $R(n)_1$ (see
Eq.~(\ref{eq-nsrep})) will contribute.
 
In Eqs.~(\ref{eq-ijk})--(\ref{eq-klm}) we have expressed the full and
the product spectra as sums over auxiliary diagrams.  Each of the
three sums contains the same {\em number\/} of auxiliary diagrams
(even though they are not necessarily the same diagrams).  One finds
that each auxiliary diagram contributes precisely one doubly atypical
superrepresentation (with $F$ columns).  Thus all three spectra
contain the same number of doubly atypical states.  In other words,
the BPS part of the spectra matches exactly.

\subsection{Explicit evaluation and comparison}
\label{sec-crank}

We now turn to checking whether the inclusion relation
Eq.~(\ref{eq-qf}) holds for the complete spectra (not just the BPS
states).  As discussed in Sec.~\ref{sec-reduce}, it will suffice to
compare $S''(m)_F$ and $S^{\rm prod}(m)_F$.  Note that the two spectra
differ solely in whether $m$ limits the number of columns, or rows, of
the auxiliary diagrams.  If the diagrams contain fewer than $m$ boxes
($F\leq m$), neither of these restrictions actually eliminate any
diagrams.  It follows immediately that the spectra agree identically
as long as the total excitation number does not exceed the number of
membranes:
\begin{equation}
S''(m)_F = S^{\rm prod}(m)_F~~~{\rm for~all}~~F\leq m~.
\label{eq-fm}
\end{equation}
This is important because it means that to verify Eq.~(\ref{eq-q}) for
all $m$ at fixed $F$, it suffices to check only a finite set of
spectra ($1\leq m\leq F-1$).\footnote{In fact it suffices to consider
  $2\leq m\leq F-3$, which follows from the trivial agreement of the
  $m=1$ spectra and by analyzing in detail the cancellation of
  restricted partitions $P(F/m)$ against $P(F\backslash m)$ for
  $m=F-2$, m=$F-1$, using Eqs.~(\ref{eq-trade1}) and (\ref{eq-trade2})
  below.}

We have used the LiE package~\cite{LiE} to expand both $S''(m)_F$ and
$S^{\rm prod}(m)_F$ into a sum of irreducible representations, as in
Eq.~(\ref{eq-expand}), for all values of the total occupation number
$F$ up to 23.  We found that Eq.~(\ref{eq-coeff}) holds.  This shows
that $S''(m)_F\supseteq S^{\rm prod}(m)_F$ for $F\leq 23$ and all
$m$.  This is evidence that $S''\supseteq S^{\rm prod}$ for all $m$,
as we expect physically.  It would be nice to prove this rigorously.

The coefficients $c(m)_F$ are listed in Appendix~\ref{sec-coeff}.
However, since we have not been able to prove our result generally, it
may be instructive to go through a few examples quite explicitly.
This will illustrate the nature of the evidence we provide, and what
remains to be proven.

Let us begin with $F=1$.  This case is trivial for any $m\geq 1$: on
both sides, there is only one auxiliary Young diagram, $\tableau{1}$,
leading to the same $U(\infty)$ representation, $\tableau{1}$.
For $F=2$, $m=1$, there is again one allowed Young diagram on each
side: $\tableau{1 1}$ for $S''$, and $\tableau{2}$ for $S^{\rm prod}$.
Hence, $S''(1)_2 = \mathsf{sym}~\tableau{1}^2 = \tableau{2}= S^{\rm
  prod}(1)_2$.  For any $m\geq 2$, each side contains the same two
auxiliary diagrams: $\tableau{2}, \tableau{1 1}$.

Next consider $F=3$.  For $m=1$ the auxiliary diagrams are $\tableau{1
  1 1}$ ($S''$) and $\tableau{3}$ ($S^{\rm prod}$).  Thus $S''(1)_3 =
\tableau{3} = S^{\rm prod}(1)_3$.  The reader will have noticed that
independently of $F$, at $m=1$ each side contains only one diagram,
and that both lead to one copy of the totally symmetrized $U(\infty)$
representation.  This is as it should: at $m=1$ the spectra $S'$,
$S''$, and $S^{\rm prod}$ must all agree, since they describe the same
object: a single membrane.  Hence we will no longer consider $m=1$
below.  At $F=3$, $m=2$, the same new auxiliary diagram appears on
both sides ($\tableau{2 1}$), and the agreement persists for larger
$m$ according to Eq.~(\ref{eq-fm}).

Thus, up to occupation number 3 the two spectra agree exactly for all
values of $m$:
\begin{equation}
S''(m)_F=S^{\rm prod}(m)_F,~~~F=1,\ 2,\ 3~.
\end{equation}

The first difference arises at $F=4$.  At $m=1$ one has the usual
identity, but $m=2$ permits the following additional auxiliary
diagrams:
\begin{equation}
  \tableau{2 1 1}~,~ \tableau{2 2} 
\end{equation}
for $S''$ and
\begin{equation}
  \tableau{3 1}~,~ \tableau{2 2}
\end{equation}
for $S^{\rm prod}$.  One can cancel the $\tableau{2 2}$ diagram since
it will contribute the same irreducible representations to both
spectra. The $\tableau{2 1 1}$ diagram yields
\begin{equation}
\tableau{4} \oplus \tableau{3 1} \oplus \tableau{2 2}~,
\end{equation}
which is a proper superset of the representation
\begin{equation}
\tableau{4} \oplus \tableau{3 1}
\end{equation}
coming from the $\tableau{3 1}$ auxiliary diagram.  For any $m>2$ one
finds again that the two spectra agree exactly.

Finally, let us go through the $F=5$ case.  In the following table, we
show for each value of $m$ (first column) the auxiliary diagrams
relevant for the full spectrum (second column) and the representations
they generate (third column) and similarly for the product spectrum
(fourth and fifth columns).  The sixth column shows the running 
difference between the previous two for each $m$; note that it always 
positive as required for the inclusion relation (\ref{eq-qu}) to hold.  
In Appendix~\ref{sec-coeff}, we continue the comparison for 
$6\leq F\leq 10$.

\begin{center}
\begin{tabular}{c||c|p{1.4in}||c|p{1.4in}||c|}
$m$ & Aux Full & $S''(m)_F$ & Aux Prod & $S^{\rm prod}(m)_F$ & $S''-S^{\rm prod}$\\
\hline
\hline
1 & $\tableau{1 1 1 1 1}$
& $\mathsf{sym}~\tableau{1}^5 = \tableau{5}$
& $\tableau{5}$
& $\mathsf{sym}~\tableau{1}^5 = \tableau{5}$
& 0 \\
\hline
\hline
2 & $\tableau{2 1 1 1}$
& $\mathsf{sym}~ \tableau{1}^3 \otimes \mathsf{sym}~ \tableau{1}^2 
   = \tableau{5} \oplus \tableau{4 1} \oplus \tableau{3 2}$
& $\tableau{4 1}$
& $\mathsf{sym}~ \tableau{1}^4 \otimes \tableau{1} 
   = \tableau{5} \oplus \tableau{4 1}$
& \\
\cline{2-5}
& $\tableau{2 2 1}$
& $\mathsf{sym} \! \left(\mathsf{sym}~\tableau{1}^2\right)^2 \otimes \tableau{1} 
   = \tableau{5} \oplus \tableau{4 1} \oplus \tableau{3 2} \oplus \tableau{2 2 1}$
& $\tableau{3 2}$
& $\mathsf{sym}~ \tableau{1}^3 \otimes \mathsf{sym}~ \tableau{1}^2
   = \tableau{5} \oplus \tableau{4 1} \oplus \tableau{3 2}$
& $\tableau{3 2} \oplus \tableau{2 2 1}$ \\
\hline
\hline
3 & $\tableau{3 1 1}$
& $\mathsf{sym}~\tableau{1}^3 \otimes \mathsf{sym}~ \tableau{1}^2
   = \tableau{5} \oplus \tableau{4 1} \oplus \tableau{3 2}$
& $\tableau{3 1 1}$
& $\mathsf{sym}~\tableau{1}^3 \otimes \mathsf{sym}~ \tableau{1}^2
   = \tableau{5} \oplus \tableau{4 1} \oplus \tableau{3 2}$
& \\
\cline{2-5}
 & $\tableau{3 2}$
& $\mathsf{sym}~ \tableau{1}^3 \otimes \mathsf{sym}~ \tableau{1}^2
   = \tableau{5} \oplus \tableau{4 1} \oplus \tableau{3 2}$
& $\tableau{2 2 1}$
& $\mathsf{sym} \! \left(\mathsf{sym}~\tableau{1}^2\right)^2 \otimes \tableau{1}
   = \tableau{5} \oplus \tableau{4 1} \oplus \tableau{3 2} \oplus \tableau{2 2 1}$
& $\tableau{3 2}$ \\
\hline
\hline
4 & $\tableau{4 1}$
& $\mathsf{sym}~ \tableau{1}^4 \otimes \tableau{1}
   = \tableau{5} \oplus \tableau{4 1}$
& $\tableau{2 1 1 1}$
& $\mathsf{sym}~ \tableau{1}^2 \otimes \mathsf{sym}~ \tableau{1}^3
   = \tableau{5} \oplus \tableau{4 1} \oplus \tableau{3 2}$
& 0 \\
\hline
\hline
5 & $\tableau{5}$
& $\mathsf{sym}~\tableau{1}^5 = \tableau{5}$
& $\tableau{1 1 1 1 1}$
& $\mathsf{sym}~\tableau{1}^5 = \tableau{5}$
& 0 \\
\hline
\end{tabular}
\end{center}

\section{Mapping states}
\label{sec-map}

\subsection{Conjugation of auxiliary diagrams}
\label{sec-conjugate}

In the previous section we have found evidence that the full spectrum
of $m$ coincident membranes contains all representations expected from
the product states of the individual membranes (and more).  However,
we only counted multiplicities of representations; we did not ask how
a given product state should be mapped into a particular state in the
full spectrum.  In this section we will learn about some aspects of
this mapping.  For simplicity, we take $n=1$; then the single particle
spectrum contains only one irreducible representation of the symmetry
group.

It is instructive to begin by considering an oversimplified case.  Let
us suppose that the symmetry group is $U(1)$ and the one-particle
spectrum consists only of the fundamental representation of $U(1)$.
(This would correspond to a bosonic model with only one transverse
direction of oscillation.)  In this case every auxiliary Young diagram
with $g$ boxes leads to the same representation: the totally
symmetrized representation with $g$ indices, containing only one
state.  The product and full spectra then match exactly, since they
contain the same number of auxiliary diagrams.\footnote{The same
  mapping works for the BPS-protected states~\cite{DSV1,DSV2} in the
  real plane wave matrix model (see Sec.~\ref{sec-bps}).  In both
  cases, there is only one relevant representation for each auxiliary
  Young diagram.}

One might expect that for more general representations it should also
be possible to find a mapping of entire auxiliary Young diagrams of
the product spectrum into auxiliary Young diagrams of the full
spectrum.  The only natural way to do this would be conjugation, the
map that flips the Young diagram about its diagonal axis.

However, explicit counterexamples show that this naive mapping cannot
be correct.  For example, the auxiliary diagram $\tableau{4 3 3}$ in
the $F=10$, $m=3$ product spectrum contains $U(\infty)$
representations which are not generated by the $\tableau{3 3 3 1}$
diagram of the full spectrum.  It is only after summing over all
allowed auxiliary diagrams at each value of $m$ that the full spectrum
becomes a superset of the product spectrum.  This suggests that the
true map is quite scrambled.

Interestingly, an inclusion relation does hold at the level of
individual auxiliary diagrams between products of {\em
  distinguishable\/} particles and traces.  This is proven in
Appendix~\ref{sec-dist}.

\subsection{Center of mass ground states}
\label{sec-com}

One can learn more about the details of the mapping by considering
states in which the center of mass degree of freedom is not excited
(``c.o.m.\ ground states'' for short).  It is straightforward to
construct all such states in the full spectrum~\cite{DSV2}.  Let us
denote the $m\times m$ matrix degrees of freedom (such as $X$, $Y$,
and $\psi$ in the $SU(4|2)$ model) collectively by $A$, and define
\begin{equation}
B = A- \frac{{\rm Tr}~A}{m}~\mathbf{1}~.
\end{equation}
The trace of $B$ vanishes but traces of higher powers of $B$ (up to
$m$) are nonzero and mutually independent.

Using $B$ instead of $A$, c.o.m.\ ground states are constructed as in
Sec.~\ref{sec-finite}, except that there will be no contributions
from traces of only one matrix.  This corresponds to requiring that
the first and second column of the auxiliary Young diagram have the
same height.  All other diagrams vanish because they contain powers of
${\rm Tr} B=0$.

On the other hand, product states without center of mass excitations
are annihilated by all center of mass annihilation operators:
\begin{equation}
  a_{\rm com}|\psi\rangle = 0~,
\label{eq-ac}
\end{equation}
where $a_{\rm com}=m^{-1/2} \sum_{i=1}^m a_i$.  Here the index $i$
refers to the $i$-th ``membrane'', and Eq.~(\ref{eq-ac}) must hold for
every operator in the single-particle multiplet.

Now consider the simple example $F=2$, $m=2$, with symmetry group
$U(1)$.  (Recall that with $U(1)$ symmetry, both auxiliary diagrams
yield only one representation, $\tableau{2}$.)  The full spectrum has
precisely one center of mass ground state, coming from the
$\tableau{2}$ auxiliary diagram for the $B$ operators.  

In the product spectrum, we again represent each state by the
corresponding auxiliary diagram.  The center of mass annihilation
operator acts as follows:
\begin{equation}
a_{\rm com}\tableau{2} = \tableau{1} 
\end{equation}
and also
\begin{equation}
a_{\rm com}\tableau{1 1} = \tableau{1} ~.
\end{equation}
Hence the product state $\tableau{2}-\tableau{1 1}$ contains no center
of mass energy.  

On both sides, we have found the same number (1) of c.o.m.\ ground
states, as expected in the $U(1)$ case where the spectra should match
exactly.  But the linear combination of product-spectrum auxiliary
diagrams differs from that for the $B$ operators ($\tableau{2}$) and
also from that of the original $A$ operators
[$\tableau{2}-\frac{1}{2}\tableau{1 1}$, since ${\rm Tr} B^2 = {\rm
  Tr} A^2 - \frac{1}{m} ({\rm Tr} A)^2]$ of the full spectrum.

One feature of this example turns out to generalize to arbitrary $F$
and $m$ (a proof is given in Appendix~\ref{sec-comproof}): Each
c.o.m.\ ground state in the product spectrum is a linear combination
containing precisely one auxiliary diagram whose first two rows have
equal length.  These are the diagrams conjugate to the auxiliary
diagrams defining the c.o.m.\ ground states of the full spectrum,
which demonstrates that conjugation of Young diagrams does play a role
in the mapping.

However, we conclude from the above example that even for the simple
case of a $U(1)$ symmetry group, where the number of states manifestly
agree on both sides, an individual product state is generically mapped
into a linear combination of states in the natural basis for the full
spectrum.  It would be nice to understand this nontrivial mapping
better.

\section{M-theory limit}
\label{sec-mlimit}

So far we have been working in the limit $\mu\to\infty$ at fixed
arbitrary $N$.  The M-theory limit corresponds to taking $N\to\infty$
at fixed $\mu$ and fixed longitudinal momentum $N/R$.  We have
focussed on vacua with $m$ coincident membranes, corresponding to the
partition $(n^m)$.  In this case the M-theory limit is $n\to\infty$
while holding $m$ and $n/R$ fixed.

As mentioned in Sec.~\ref{sec-free}, the effective coupling in the
coincident membrane configuration is
\begin{equation}
\lambda = \left(\frac{R}{\mu n}\right)^3 m~,
\end{equation}
which is fixed in the M-theory limit, suggesting that perturbation
theory remains valid~\cite{DSV1}.

On the other hand, Ref.~\cite{DSV1} identified an explicit path
leading to a change in the membrane configuration, with barrier height
of order 
\begin{equation}
\frac{1}{n} \left(\frac{\mu n}{R}\right)^3~.
\label{eq-path}
\end{equation}
(A membrane following this path extends a spike inward to its center
and grows a small new spherical membrane there.)  This barrier height
vanishes like $1/n$ in the M-theory limit, suggesting that
perturbation theory breaks down.  

This apparent paradox prompted the speculation~\cite{DSV1} that the
low energy decay path is very narrow in the space of perturbations
about the minimum, and that thus perturbation theory is valid for {\em
  generic\/} perturbations up to much higher energies than
(\ref{eq-path}).

Here we note that $\mu$ is no longer a physical parameter after
$N\to\infty$ has been taken.  It can be set to any value by a change
of coordinates.  Hence the question of whether or not a vacuum is
perturbative in the M-theory limit cannot depend on the value of
$\mu$.  But taking the M-theory limit at small $\mu$ suggests that
perturbation theory is completely invalid even for generic
perturbations.  This apparently conflicts with the expectation of a
perturbative regime if we first take the M-theory limit at large $\mu$
and then rescale coordinates.

We expect that these two pictures are reconciled as follows.  In the
M-theory limit at large $\mu$ there is a small but finite decay rate
for any vacuum at any finite temperature $T>0$.  It is suppressed not
by an energy barrier but by the time it takes to find the low pass
through a high mountain range.  Physically the coordinate rescaling of
$\mu$ corresponds to a boost, and rescaling from very large to small
$\mu$ requires an enormous boost, or more precisely, an undoing of the
original boost that made $\mu$ large.  The unboosted observer will
find that all processes happen at a much higher rate, so the decay and
mixing of vacua will be seen to occur instantaneously.

We hope to return to this question in more detail in future work.

\acknowledgments

We are grateful to T.~Banks, D.~Berenstein, P.~Ho\v{r}ava, C.~Keeler,
J.~Maldacena, A.~Pasqua, S.~Shenker, W.~Taylor, and B.~Tweedie for
helpful discussions and comments.  We are especially indebted to
M.~Aganagic and M.~Van Raamsdonk for detailed explanations.

\appendix

\section{Coefficients of irreducible representations}
\label{sec-coeff}

In this appendix we list explicit results of a comparison between the
full spectrum and the product spectrum.  We have carried out this
check for values up to and including $F=23$.  The computational
complexity and memory requirements grow exponentially with $F$.  In
all cases we have found that the product states form a subset of the
full set of representations.  This shows up in the table below in that
all multiplicities of representations in the difference are positive.

Due to space constraints we list explicit results only for $6 \leq F
\leq 10$ and all relevant $m$.  The comparison for smaller $F$ was
already carried out in Sec.~\ref{sec-crank}.  As explained above, for
$m=1$ and $m \geq F$, the irreps on both sides match exactly, giving
no net difference for these values of $m$.  Therefore, in the table
below, we do not go beyond $m=F-1$ (but include $m=1$ for
completeness).  

\begin{center}
\tablefirsthead{%
}
\tablehead{%
  \hline
  & \multicolumn{2}{|l|}{\small\sl continued from previous page}\\
  \hline}
\tabletail{%
  \hline
  & \multicolumn{2}{|r|}{\small\sl continued on next page}\\
  \hline}
\tablelasttail{%
}
\begin{supertabular}{c||c||p{5.2in}|}
$F$ & $m$ & $S''(m)_F-S^{\rm prod}(m)_F$ \\
\hline
\hline
6 & 1 & -- \\
\cline{2-3}
  & 2 & $2(4~2) + (3~2~1) +2(2^3)$ \\
\cline{2-3}
  & 3 & $2(4~2) + (3~2~1) +(3^2) + (2^3)$ \\
\cline{2-3}
  & 4 & $(4~2)$ \\
\cline{2-3}
  & 5 & -- \\
\hline
\hline
7 & 1 & -- \\
\cline{2-3}
  & 2 & $2(5~2)+(4~3)+2(4~2~1)+2(3~2^2)+(2^3~1)$ \\
\cline{2-3}
  & 3 & $3(5~2)+3(4~3)+3(4~2~1)+(3^2~1)+3(3~2^2)+(2^3~1)$ \\
\cline{2-3}
  & 4 & $2(5~2)+2(4~3)+(4~2~1)+(3~2^2)$ \\
\cline{2-3}
  & 5 & $(5~2)$\\
\cline{2-3}
  & 6 & -- \\
\hline
\hline
8 & 1 & -- \\
\cline{2-3}
  & 2 & $3(6~2)+(5~3)+2(5~2~1)+2(4^2)+(4~3~1)+4(4~2^2)
         +(3~2^2~1)+(2^4)$\\
\cline{2-3}
  & 3 & $5(6~2)+5(5~3)+5(5~2~1)+3(4^2)+5(4~3~1)+7(4~2^2)
         +(4~2~1^2)+2(3^2~2)+2(3~2^2~1)+2(2^4)$ \\
\cline{2-3}
  & 4 & $4(6~2)+4(5~3)+3(5~2~1)+4(4^2)+3(4~3~1)+4(4~2^2)
         +(3^2~2)+(3~2^2~1)+(2^4)$ \\
\cline{2-3}
  & 5 & $2(6~2)+2(5~3)+(5~2~1)+(4^2)+(4~2^2)$ \\
\cline{2-3}
  & 6 & $(6~2)$ \\
\cline{2-3}
  & 7 &  -- \\
\hline
\hline
9 & 1 & -- \\
\cline{2-3}
  & 2 & $3(7~2)+2(6~3)+3(6~2~1)+2(5~4)+(5~3~1)+4(5~2^2)+2(4^2~1)
         +2(4~3~2)+2(4~2^2~1)+2(3~2^3)+(2^4~1)$ \\
\cline{2-3}
  & 3 & $7(7~2)+9(6~3)+8(6~2~1)+6(5~4)+9(5~3~1)+12(5~2^2)
         +2(5~2~1^2)+7(4^2~1)+9(4~3~2)+2(4~3~1^2)+6(4~2^2~1)
         +(3^2~2~1)+4(3~2^3)+(2^4~1)$ \\
\cline{2-3}
  & 4 & $6(7~2)+9(6~3)+7(6~2~1)+8(5~4)+8(5~3~1)+9(5~2^2)+(5~2~1^2)
         +7(4^2~1)+8(4~3~2)+(4~3~1^2)+4(4~2^2~1)+(3^2~2~1)
         +3(3~2^3)+(2^4~1)$ \\
\cline{2-3}
  & 5 & $4(7~2)+5(6~3)+3(6~2~1)+5(5~4)+3(5~3~1)+4(5~2^2)+2(4^2~1)
         +2(4~3~2)+(4~2^2~1)+(3~2^3)$ \\
\cline{2-3}
  & 6 & $2(7~2)+2(6~3)+(6~2~1)+(5~4)+(5~2^2)$ \\
\cline{2-3}
  & 7 & $(7~2)$ \\
\cline{2-3}
  & 8 & -- \\
\hline
\hline
10 & 1 & -- \\
\cline{2-3}
   & 2 & $4(8~2)+2(7~3)+3(7~2~1)+4(6~4)+2(6~3~1)+6(6~2^2)
          +2(5~4~1)+2(5~3~2)+2(5~2^2~1)+4(4^2~2)+(4~3~2~1)
          +4(4~2^3)+(3~2^3~1)+2(2^5)$ \\
\cline{2-3}
   & 3 & $9(8~2)+13(7~3)+12(7~2~1)+13(6~4)+16(6~3~1)+19(6~2^2)
          +3(6~2~1^2)+3(5^2)+14(5~4~1)+18(5~3~2)+5(5~3~1^2)
          +(5~2^2~1)+13(4^2~2)+4(4^2~1^2)+3(4~3^2)+8(4~3~2~1)
          +9(4~2^3)+(4~2^2~1^2)+2(3^2~2^2)+2(3~2^3~1)+2(2^5)$ \\
\cline{2-3}
   & 4 & $10(8~2)+15(7~3)+12(7~2~1)+19(6~4)+19(6~3~1)+19(6~2^2)
          +3(6~2~1^2)+4(5^2)+18(5~4~1)+20(5~3~2)+5(5~3~1^2)
          +10(5~2^2~1)+16(4^2~2)+3(4^2~1^2)+4(4~3^2)+8(4~3~2~1)
          +9(4~2^3)+(4~2^2~1^2)+2(3^2~2^2)+2(3~2^3~1)+2(2^5)$ \\
\cline{2-3}
   & 5 & $7(8~2)+10(7~3)+7(7~2~1)+13(6~4)+10(6~3~1)+10(6~2^2)
          +(6~2~1^2)+5(5^2)+10(5~4~1)+10(5~3~2)+(5~3~1^2)
          +4(5~2^2~1)+7(4^2~2)+4(4~2^3)+(4^2~1^2)+(4~3^2)
          +3(4~3~2~1)+(3^2~2^2)+(3~2^3~1)+(2^5)$ \\
\cline{2-3}
   & 6 & $4(8~2)+5(7~3)+3(7~2~1)+6(6~4)+3(6~3~1)+4(6~2^2)+(5^2)
          +2(5~4~1)+2(5~3~2)+(5~2^2~1)+(4^2~2)+(4~2^3)$ \\
\cline{2-3}
   & 7 & $2(8~2)+2(7~3)+(7~2~1)+(6~4)+(6~2^2)$ \\
\cline{2-3}
   & 8 & $(8~2)$ \\
\cline{2-3}
   & 9 & -- \\
\hline
\end{supertabular}
\end{center}

\section{Representations and plethysm: Notation, examples, and
  properties}
\label{sec-math}

In this appendix we summarize various definitions and properties
concerning group representations and operations between them.  Further
details and references can be found, e.g., in
Ref.~\cite{FauJar05}.

\subsection{Young diagrams and tableaux}

The irreducible tensor representations of $U(m)$ with $r$ indices
correspond to all possible ways of (anti-)symmetrizing the $r$-th power
of the fundamental representation.  Hence, they are in one-to-one
correspondence with the representations of the symmetric group, and
they can be denoted by Young diagrams with $r$ boxes and up to $m$
rows.  These diagrams also label the partitions of $r$ with up to $m$
positive integer terms, where each row represents one term.\footnote{%
  Alternatively, these diagrams also label the partitions with no term
  greater than $m$, where each column represents one term.}

We denote an irreducible representation $A$ by its Young diagram,
(e.g., $A=\tableau{4 2 2}$) or by the corresponding partition of $r$,
in the standard notation where ``exponents'' denote repeated terms
(e.g., $A=(4~2^2)$).  We denote the sum of two representations by $A
\oplus B$ and the tensor product by $A\cdot B$.  For $A\cdot A$ we
also write $A^2$.  For example, in $U(3)$,
\begin{equation}
  \tableau{1 1}^2 \equiv \tableau{1 1} \cdot \tableau{1 1} =  
  \tableau{2 1 1}~\oplus~\tableau{2 2}~.
\label{eq-11sq}
\end{equation}

A Young tableau is a Young diagram filled out with positive integers
no larger than $m$, such that the numbers in each row are
nondecreasing from left to right, and the numbers in each column are
increasing from top to bottom.  For a given Young diagram, the number
of such tableaux is given by the usual ``hook rule'' and is equal to
the dimension of the corresponding representation.  (The requirements
cannot be satisfied for diagrams with more than $m$ rows, which is why
they have no states and can be ignored.)

\subsection{Schur functions}

Another way to describe an irreducible representation $A$ is to write
down its Schur function $s_{A}$.  The Schur function is the character
(i.e., the trace) of the matrix representing a group element, as a
function of (the conjugacy class of) the group element.  It is a
function of $m$ variables.

The Schur function is easily be obtained from the Young diagram as
follows.  For each allowed tableau, write down a product of $x$'s
indexed with the numbers appearing in the tableau; then add them all
up.  This will produce a symmetric polynomial of degree $r$ with
positive coefficients whose sum is the dimension.  For example, for
$U(3)$, the representation $\tableau{2 1}$ (dimension 8) has the Schur
function
\begin{equation}
  s_{\tableau{2 1}}(x_1,x_2,x_3) = 
  x_1^2 x_2 + x_1^2 x_3 + x_2^2 x_3 + x_2^2 x_1 + x_3^2 x_1 + x_3^2 x_2
  + 2x_1x_2x_3~.
\label{eq-s21}
\end{equation}

Schur functions of irreducible representations form a basis for the
symmetric polynomials.  The Schur functions of a reducible
representation is the sum of the Schur functions of the irreducible
representations it contains.  One way to decompose the product of two
representations is to multiply their Schur functions and expand the
result in the basis of Schur functions of the appropriate degree.
This reproduces the usual Littlewood-Richardson rule.  For example, in
$U(3)$,
\begin{eqnarray}
  \tableau{1 1}~\cdot~\tableau{1} & = & 
  (x_1 x_2 + x_1 x_3 + x_2 x_3)(x_1 + x_2 + x_3) \cr
  & = & x_1^2 x_2 + x_1^2 x_3 + x_2^2 x_3 + 
  x_2^2 x_1 + x_3^2 x_1 + x_3^2 x_2 + 3x_1x_2x_3 \cr
  & = & \tableau{2 1}~\oplus~\tableau{1 1 1}~.
\label{eq-s11s1}
\end{eqnarray}

If we are interested only in the final sum of diagrams, this
computation can be carried out for the lowest value of $m$ for which
all resulting representations exist (definitely $m$ need not be larger
than the total number of boxes).  The result will hold for all $m$;
for small $m$, diagrams with more than $m$ boxes can be deleted.

\subsection{Plethysm}

\paragraph{Definition} Plethysm is the composition of two Schur
functions, denoted as $A^{\otimes B}$.  Let $A$ be a representation of
$U(m)$, and let $B$ be an arbitrary representation of $U(\dim A)$.
(Neither $A$ nor $B$ are assumed irreducible.)  Write the Schur
function for $A$ as $s_A=\sum_{i=1}^{\dim A} y_i$, where $y_i$ are the
monomials arising from the individual Young tableaux.  Plethysm of $A$
by $B$ is defined as
\begin{equation}
A^{\otimes B} \equiv s_B(y)~.
\label{eq-plethdef} 
\end{equation}

For example, for the $U(3)$ representation $A=\tableau{1 1}$ one has
$y_1=x_1 x_2$, $y_2=x_1 x_3$, and $y_3=x_2 x_3$.  Let us take
$B=\tableau{1 1}$.  Then
\begin{eqnarray} 
\tableau{1 1}^{\otimes\tableau{1 1}}
 & = & y_1 y_2 + y_1 y_3 + y_2 y_3\cr
 & = & x_1^2 x_2x_3 + x_3^2 x_1x_2 + x_2^2 x_1x_3\cr
 & = & \tableau{2 1 1}~ .
\label{eq-11p11}
\end{eqnarray} 

\paragraph{Properties} Plethysm has a number of properties that are
used in this paper.  It is clear from the definition that plethysm is
{\em right distributive\/}, i.e., distributive in $B$:
\begin{equation}
  A^{\otimes (B_1\oplus B_2)}=A^{B_1}\oplus A^{B_2}~.
\end{equation}
It is also {\em associative\/}:
\begin{equation}
\left(A^{\otimes B}\right)^{\otimes C} = A^{\otimes\left(B^{\otimes C}\right)}~.
\end{equation}
The $g$-th tensor power of $A$ can be written as the sum over
all plethysms of $A$ by Young diagrams with $g$ boxes:
\begin{equation}
A^g = \bigoplus_{P(g)}  A^{\otimes P}~,
\end{equation}
where the sum is over all partitions of $g$.  [Comparison of
Eqs.~(\ref{eq-11p11}) and (\ref{eq-11p2}) with Eq.~(\ref{eq-11sq})
will give an example of this for $A=\tableau{1 1}$ and $g=2$.]

\paragraph{Special cases} In this paper we frequently consider two
special cases of plethysm which get their own name.  The first is
$B=(g)$, i.e., the Young diagram with one row, corresponding to the
totally symmetric representation with $g$ indices.  We denote plethysm
by $(g)$ as $\mathsf{sym}~ A^g$:
\begin{equation}
\mathsf{sym}~A^g \equiv A^{(g)}~.
\label{eq-symdef}
\end{equation}
E.g., for the $U(3)$ representation $A=\tableau{1 1}$
\begin{eqnarray} 
  \mathsf{sym}~ A^2 \equiv
  \tableau{1 1}^{\otimes\tableau{2}}
  & = & y_1^2+y_2^2+y_3^2+y_1 y_2 + y_1 y_3 + y_2 y_3\cr
  & = & x_1^2x_2^2+x_1^2x_3^2+x_2^2x_3^2
  +x_1^2 x_2x_3 + x_3^2 x_1x_2 + x_2^2 x_1x_3\cr
  & = & \tableau{2 2}~ .
\label{eq-11p2}
\end{eqnarray} 

The second important case is $B=({\rm oddrow}_g)$, i.e., the sum of
all Young diagrams with $g$ boxes and an odd number of (nonzero)
rows.  We denote plethysm by $({\rm oddrow}_g)$ as $\mathsf{cyc}~
A^g$:
\begin{equation}
\mathsf{cyc}~A^g \equiv A^{\otimes ({\rm oddrow}_g)}~.
\label{eq-cycdef}
\end{equation}
E.g., for the $U(3)$ representation $A=\tableau{1 1}$
\begin{eqnarray} 
  \mathsf{cyc}~ A^3 \equiv
  \tableau{1 1}^{\otimes(\tableau{3}\oplus\tableau{1 1 1})}
  = \tableau{1 1}^{\otimes\tableau{3}} \oplus \tableau{1 1}^{\otimes\tableau{1 1 1}}
  & = & \tableau{3 3} \oplus \tableau{2 2 2}~ .
\label{eq-11c3}
\end{eqnarray} 

\paragraph{Interpretation} Plethysm is interpreted as follows.  Think
of $B$ as a representation of the permutation group, and let $g$ be
the number of boxes in the Young diagram of $B$.  Plethysm of $A$ by
$B$ computes the restriction of the $g$-th tensor power of $A$ to the
symmetrization pattern $B$.  

Let us discuss this in more detail for the two types of plethysm
mainly considered in this paper.  Suppose $A$ represents a $U(m)$
multiplet of creation operators (i.e., $A$ carries $U(m)$ indices but
it is not a matrix). If all creation operators commute, the set of
states created by $AAA|0\rangle$ will form the representations
computed by $\mathsf{sym}~A^3 \equiv A^{\otimes\tableau{3}}$, and {\em
  not by $A^3$}, since exchange of the complete set of indices between
different $A$'s gives the same state.

The $\mathsf{cyc}$ plethysm defined in Eq.~(\ref{eq-cycdef}) computes
the representations formed by the trace of a product of $g$ matrices
of rank $h\geq g$:
\begin{equation}
{\rm Tr} AAA\ldots A|0\rangle~,
\label{eq-tra}
\end{equation}
where the matrix $A$ represents a $U(m)$ multiplet of matrix
operators.  

To see this, note that the correct plethysm $B$ will be the sum of all
$g$-box Young diagrams $B_{\alpha}$ whose symmetrization pattern is
compatible with the manifest invariance of Eq.~(\ref{eq-tra}) under
cyclic exchange of indices:
\begin{equation}
B=\sum_{\alpha} B_\alpha~.
\end{equation}
It remains to show that a $g$-box Young diagram satisfies this
criterion if and only if it has an odd number of rows.

Consider filling a given Young diagram $B_\alpha$ with integers
$1\ldots g$ from left to right, top to bottom row.  An elementary
cyclic permutation produces a different Young tableau by shifting
every integer to the right, and the integer at the end of each row to
the beginning of the next row; $g$ now occupies the top left box.
Invariance under cyclic permutations requires that these two Young
tableaux appear with the same sign in the symmetrization pattern
$B_\alpha$.  Recall that this consists of first totally symmetrizing
within each row, then totally antisymmetrizing over the indices
appearing within each column.  Starting from the first Young tableau,
we rearrange the entries in each row so that they agree with the
second Young tableau everywhere except in the first column.  Since
this involves only manipulations within rows we pick up no minus sign.
To end up with the second tableau, we now need to move all entries in
the first row down by one box (and the $g$ up to the top left box).
If $B_\alpha$ has $q$ rows, this will involve $q-1$ pairwise
permutations.  Each contributes a factor $(-1)$, so the overall sign
is $(-1)^{q-1}$, which is positive precisely if $q$ is odd.  This
shows that odd $q$ is necessary.  It is also sufficient, since all
other cyclic permutations are powers of the elementary one we
considered.

\section{Trace relations}
\label{sec-rel}

In this appendix we discuss the efficacy of the truncation in
Eq.~(\ref{eq-fullsp}) in eliminating redundant states.  We consider a
general case with $k$ different matrices of size $m\times m$.  In the
context of this paper, $k=$ dim$(R(n)_1)$; see Eq.~(\ref{eq-r1}).

The case of only a single species, $k=1$, is well understood.  Taking
into account gauge invariance under $U(m)$, there are $m$ physical
degrees of freedom.  Single trace states ${\rm Tr} (A^i) |0\rangle$ of
length up to $m$ are mutually independent.  For $i>m$ the trace is
always equal to a product of shorter traces.  All independent
multitrace states are thus constructed from single traces of length up
to $m$.

\subsection{Algebraic relations}

For more than one species of matrix creation operator, $k>1$, the
situation is more complicated.\footnote{The following argument was
  explained to us by M.~Van Raamsdonk~\cite{VanU}.} Gauge invariance
allows the elimination of up to $m^2$ gauge degrees of freedom,
leaving $(k-1)m^2$ physical degrees of freedom.  But there are at
least $k^F/F$ single traces of length $F$ that can be written down,
(the division by $F$ taking into account the cyclic identity of the
trace).  This number grows quickly with $F$.  It is easy to see that
for large enough $m$, $(k-1)m^2<k^F/F$ for values of $F$ at least of
order $\frac{\log m}{\log k}$, which is much smaller than $m$.

Let us imagine for a moment that the matrices consist not of operators
but of numbers.  Taking a trace does not increase the number of
independent parameters, so most of the traces of length greater than
$\frac{\log m}{\log k}$ must be algebraically related.  This means
that most traces of length up to $m$ (which can be much larger) are
not {\em algebraically}\/ independent of each other.

However, this argument does not carry over straightforwardly to matrix
{\em operators}.  Even simple algebraic relations may involve, e.g.,
some square roots, and thus will not hold between states obtained from
matrix operators acting on the vacuum.  Hence, the above counting
argument does not automatically imply that the single-trace states of
length up to $m$ fail to be independent~\cite{VanU}.  (In fact we will
argue below that they are indeed independent.)

On the other hand, it may be possible to convert some algebraic
relations into linear relations between polynomials of single traces.
This would correspond to trace relations between certain {\em
  multitrace} states.  In fact, it is possible to construct an
explicit example (with $m=2$, $F=10$, $k\geq 5$) of an identity
relating apparently different multitrace states, each of which
involves only products of traces of length up to $m$~\cite{VanU}.
This demonstrates that not all the states we consider in the spectrum
$R'$ are independent: we are overcounting states and representations
in $R'$.

We will now argue, however, that the spectrum $R'$ is an excellent
approximation to the true spectrum for values of $F\ll km^2$.  (Recall
that the total occupation number $F$ is the number of matrix creation
operators acting on the vacuum.)  Note that $k$ is of order 10 already
for $n=1$ (one unit of momentum per fuzzy sphere), grows quadratically
with $n$, and diverges in the M-theory limit.

\subsection{Thermodynamic limit}
\label{sec-thermo}

We begin with an argument explaining this scale.  For $m\to\infty$,
there are no trace relations.  In this case the number of single trace
states really is at least $k^F/F$, which gives a Hagedorn-like
spectrum.  The same will hold for multitrace states of total length
$F$ since the entropy contribution from partitioning $L$ into
individual traces is subleading.  As pointed out in
Ref.~\cite{AhaMar03}, this behavior cannot persist for finite $m$ at
energies $E\approx F$ much greater than $km^2$, because it would
contradict the thermodynamics of a finite quantum mechanical system
with $~km^2$ degrees of freedom.

The thermodynamic treatment becomes valid when the temperature is at
least of order unity.  Then $E \sim km^2 T$ and $S\sim km^2 \log T$.
Thus, for $E>km^2$, the entropy should grow only logarithmically with
the total occupation number:
\begin{equation}
S\sim km^2 \log\frac{E}{km^2}~.
\end{equation}

The only natural explanation for the discrepancy with the case of
infinite matrices, of course, is the appearance of trace relations for
finite matrices.  Hence, trace relations must become important at
energies above $F\sim km^2$.  Below this energy scale the finiteness
of the number of degrees of freedom should be difficult to detect.
The system should behave display the Hagedorn-like behavior of
infinite matrices.  Hence, one expects trace relations to have a
negligible influence on the spectrum for $F\ll km^2$.

\subsection{Exact number of states: A bosonic model}

Next, we derive a formula for the precise number of states, fully
taking into account trace relations.  We will find that no trace
relations arise for total occupation number $F\leq m$, justifying our
assumption that single trace states of length up to $m$ are mutually
independent.  For larger energies, we will compare our result to the
number of states in the approximation we employ in Sec.~\ref{sec-full}
(obtained by limiting the length of individual traces to $m$ and
ignoring trace relations).  In the regime where we can evaluate our
two expressions explicitly, their agreement will turn out to be
excellent, with no evidence for a significant effect from trace
relations.

We will carry this out for a purely bosonic model but we expect that
the trace relations will have the same (small) effect when fermions
are included.

\subsubsection{Exact number of states of the full spectrum}

Consider a theory with $k$ bosonic matrix creation operators, each of
which contributes unity to the total ``energy'' $F$.  We are
interested in the case where the operators transform in the adjoint of
$U(m)$.

The exact number of states with occupation number $F$, ${\cal N}(F)$,
is given by the coefficient of $x^F$ in the partition function, where
$x=e^{-\beta}$.\footnote{We parametrize the spectrum by total
  occupation number $F$, which is equal to the energy only if all
  operators contribute unit energy.  In the plane wave model this is
  true only approximately. Here $\beta$ is conjugate to $F$ and thus
  is only approximately an inverse temperature.}  The partition
function is given by an integral over the group\footnote{We thank
  M.~Van Raamsdonk for suggesting this method of evaluating the
  spectrum.}  $U(m)$~\cite{AhaMar03}:
\begin{equation}
  Z(x) = \int [dU] \exp \sum_{n=1}^\infty k x^n \frac{{\rm Tr} (U^n)
    {\rm Tr} (U^{\dagger n})}{n}~.
\end{equation}
Taylor expansion yields
\begin{eqnarray}
  Z(x) & = & \int [dU] \prod_{n=1}^\infty 
  \left(\sum_{q_n=0}^\infty x^{nq_n}~\frac{[({\rm Tr} (U^n)~{\rm Tr} 
      (U^{\dagger n})]^{q_n}}{n^{q_n} q_n!}
  \right)^k \\
  & = & \int [dU] \left(\sum_{\vec q\,} x^{|\vec q\,|}~\frac{{\rm
        Tr}_{\vec q\,} (U)~{\rm Tr}_{\vec q\,} 
      (U^\dagger)}{\zeta(\vec q\,)}\right)^k~,
\end{eqnarray}
where the sum is over all (infinite-dimensional) vectors $\vec q\, =
(q_1,q_2,...)$ with nonnegative integer coefficients, and we define
\begin{eqnarray}
  |\vec q\,| & \equiv & \sum_{n=1}^\infty n q_n~, \\
  {\rm Tr}_{\vec q\,} (U) & \equiv & \prod_{n=1}^\infty [{\rm Tr}
  (U^n)]^{q_n}~, \\
  \zeta(\vec q\,) & \equiv & \prod_{n=0}^\infty n^{q_n} q_n!~~~.
\end{eqnarray}
This can be further expanded to
\begin{equation}
  Z(x)  = \int [dU] \sum_{\vec q\,\,^{(1)}}\cdots \sum_{\vec q\,\,^{(k)}} 
  x^{|\vec Q\,|}~\frac{{\rm Tr}_{\vec Q\,} (U)~{\rm Tr}_{\vec Q\,} 
    (U^\dagger)}{\prod_{i=1}^k \zeta(\vec q\,^{(i)})}~,
\end{equation}
where $\vec Q\,$ is the sum of the $k$ vectors $\vec q\,^{(1)},\ldots,\vec
q^{(k)}$:
\begin{equation}
\vec Q\, \equiv \sum_{i=1}^k \vec q\,^{(i)}~.
\end{equation}

To extract the degeneracy at occupation number $F$, we restrict the
above sum to $|\vec Q\,|=F$:
\begin{equation} 
   {\cal N} (F) = \int [dU] \sum_{\vec q\,^{(i)}:~ |\vec
    Q|=F} \frac{{\rm Tr}_{\vec Q\,} (U)~{\rm Tr}_{\vec Q\,}
    (U^\dagger)}{\prod_{i=1}^k \zeta(\vec q\,^{(i)})}~.
\end{equation}
Only the denominator depends on the detailed decomposition of $\vec Q\,$
into $k$ vectors.  This part of the sum can be evaluated using the
identity
\begin{equation}
  \sum_{\vec q\,^{(i)}:~ \vec Q\,~{\rm fixed}} 
  \frac{\zeta(\vec Q\,)}{\prod_{i=1}^k \zeta(\vec q\,^{(i)})} = 
  k^{||\vec Q\,||}~,
\end{equation}
where
\begin{equation}
||\vec Q\,|| \equiv \sum_{n=1}^\infty Q_n~.
\end{equation}
This yields
\begin{equation}
   {\cal N} (F) = \int [dU] \sum_{\vec Q\,:~ |\vec
    Q|=F} k^{||\vec Q\,||} \frac{{\rm Tr}_{\vec Q\,} (U)~{\rm Tr}_{\vec Q\,}
    (U^\dagger)}{\zeta(\vec Q\,)}~.
\end{equation}

The integral can be evaluated as follows.  Each conjugacy class of the
symmetric group $S_h$ corresponds to a vector $\vec q\,$ with $|\vec
q|=h$.  (One can view $\vec q\,$ as a Young diagram.  The component
$q_i$ is the number of $i$-cycles; the number of elements in the
conjugacy class is $\zeta(\vec q\,)$.)  The group characters $\chi$ in
any two irreducible representations $R_1$ and $R_2$ with $h$ boxes
satisfy the orthogonality relations
\begin{eqnarray} 
  \sum_{\vec q\,} \frac{\chi_{R_1}(\vec q\,)\chi_{R_2}(\vec q\,)}{\zeta(\vec
    q)} & = & \delta_{R_1 R_2}~; \\
  \sum_R \frac{\chi_R(\vec q\,)\chi_R(\vec q\,')}{\zeta(\vec
    q)} & = & \delta_{\vec q\,\vec q\,'}~.
\end{eqnarray} 
In both cases the sum runs over all Young diagrams with $h$ boxes.

Similarly, the characters of the unitary group $U(m)$ satisfy
\begin{equation}
  \int [dU]~{\rm Tr}_{R_1} U~{\rm Tr}_{R_2} U^\dagger = \delta_{R_1 R_2}~,
\end{equation}
where $R_1$ and $R_2$ are arbitrary irreducible representations of
$U(m)$, and ${\rm Tr}_R U$ is the trace of the group element $U$ in
the representation $R$.  Recall that for finite $m$, the irreducible
representations of $U(m)$ are in one-to-one correspondence with Young
diagrams with up to $m$ rows.

Using these relations and the Frobenius formula it can be shown
that~\cite{AgaOog04}
\begin{equation}
{\rm Tr}_{\vec q\,} U = \sum_R \chi_R(\vec q\,) {\rm Tr}_R U~.
\end{equation}
The sum runs over all diagrams with $|\vec q\,|$ boxes, but for finite
$m$, only Young diagrams with up to $m$ rows will contribute since the
remaining representations do not exist.  It follows that
\begin{equation}
  \int [dU] {\rm Tr}_{\vec q\,} U~{\rm Tr}_{\vec q\,'} U^\dagger =
\sum_R \chi_R(\vec q\,)\chi_R(\vec q\,')~.
\end{equation}
Thus the total number of states with occupation number $F$ is
\begin{equation}
  {\cal N} (F) = 
  \sum_{\begin{array}{c}\vec Q\,:~ |\vec Q\,|=F \\  
      R:~b(R)=F,~r(R)\leq  m\end{array}}  
  [\chi_R(\vec Q\,)]^2 \frac{k^{||\vec Q\,||}}{\zeta(\vec Q\,)}~,
\label{eq-nf}
\end{equation}
where $b(R)$ and $r(R)$ denote the number of boxes and rows of $R$.

Note that for $F\leq m$, the restriction on the number of rows is
trivial and the number of states becomes independent of $m$ in this
regime.  This includes the limit of infinitely large matrices, in
which there are no trace relations.  Hence there will be no trace
relations for any value of $m\geq F$, as advertised
earlier.

As a consistency check, for $F\leq m$, Eq.~(\ref{eq-nf}) can be
evaluated using the identity~\cite{AgaOog04}
\begin{equation}
\int [dU]~{\rm Tr}_{\vec q\,} U~{\rm Tr}_{\vec q\,'} U^\dagger = 
\delta_{\vec q\,\vec q\,'} \zeta(\vec q\,)~.
\end{equation}
This yields
\begin{equation}
  {\cal N} (F) = 
  \sum_{\vec Q\,:~ |\vec Q\,|=F} k^{||\vec Q\,||}~,
\label{eq-nfmi}
\end{equation}
which agrees with the result obtained by expanding the $m\to\infty$
partition function~\cite{AhaMar03}
\begin{equation}
Z_\infty(x) = \prod_{n=1}^\infty \frac{1}{1-kx^n}~.
\end{equation}

\subsubsection{Exact number of states ignoring trace relations}

For $F\geq m$, the exact number of states, Eq.~(\ref{eq-nf}), will be
smaller than the $m\to\infty$ value, Eq.~(\ref{eq-nfmi}), reflecting the
redundancy of states due to trace relations at finite $m$.  As
explained in Sec.~\ref{sec-thermo} above, however, we expect the
approximation in Sec.~\ref{sec-full} to be quite accurate as long as
the energy does not become too large: $F\ll km^2$.  This approximation
consisted of ignoring trace relations (which is tantamount to taking
the size of the matrix to infinity) and instead removing all states
containing single traces of length greater than $m$.  (Please note
that $m$ no longer corresponds to the size of the matrix in the
analysis below!)

The exact number of states in this approximation can be obtained from
a tedious but straightforward generalization of results presented in
Ref.~\cite{AhaMar03}.  (Their most nontrivial aspect is the use of
Polya theory to take the cyclic symmetry of the trace into account.)
The partition function is given by
\begin{equation}
Z_{\rm ap}(x) = \exp \sum_{\eta=1}^\infty \psi_{km}(\eta)\frac{x^\eta}{\eta}~,
\end{equation}
where 
\begin{equation}
\psi_{km}(\eta) = \sum_{\sigma|\eta = 1}^m~~
\sum_{q|\sigma} \phi(q) K^{\sigma/q}~,
\end{equation}
and $\phi(q)$ is the number of coprimes of $q$ which are not larger
than $q$.  Notation such as $q|\sigma$ denotes integer divisors $q$ of
$\sigma$.

The number of states with total operator number $F$ is obtained by
expanding and extracting the coefficient of $x^F$:
\begin{equation} 
{\cal N}_{\rm ap}(F) = \sum_{\vec q\,:~|\vec q\,|=F}
  \frac{\prod_{\eta=1}^\infty [\psi_{km}(\eta)]^{q_\eta}}{\zeta(\vec q\,)}~.
\end{equation}

\begin{figure}[h]
\begin{center}
$\begin{array}{c@{\hspace{0.5in}}c}
\multicolumn{1}{l}{\mbox{}} &
	\multicolumn{1}{l}{\mbox{}} \\ [-0.53cm]
\epsfxsize=2.8in
\epsffile{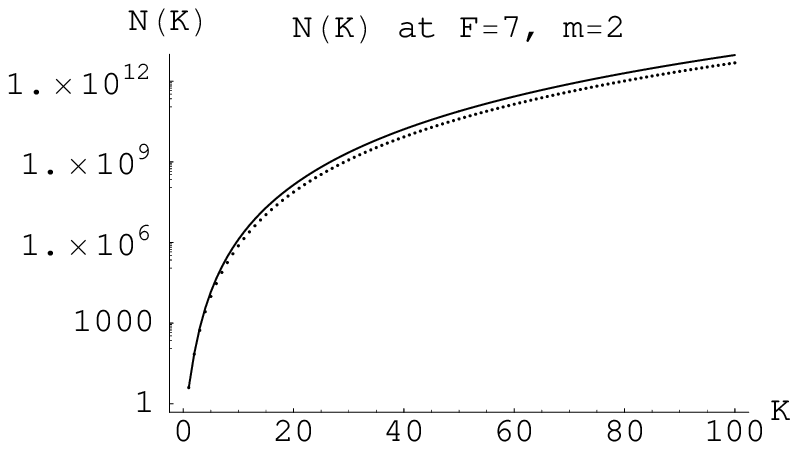} & 
\epsfxsize=2.8in
\epsffile{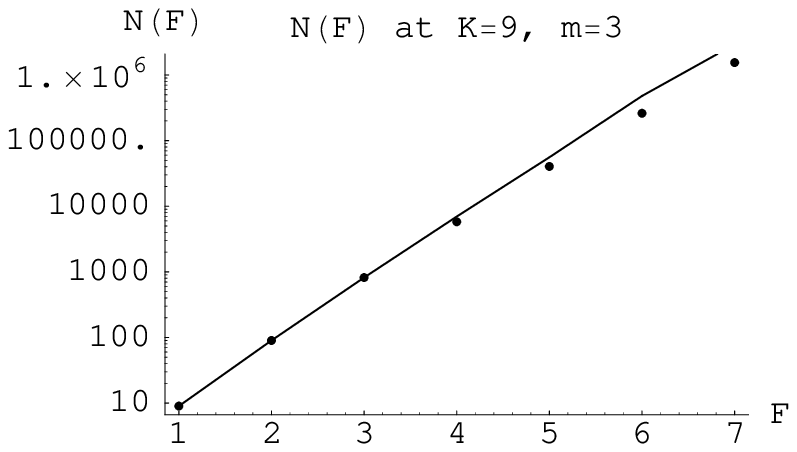}\\ [0.0cm]
\end{array}$
\end{center}
\caption{The solid line shows the exact number of states ${\cal N}$,
  taking into account trace relations.  The dots shows the number of
  states ${\cal N}_{\rm ap}$ resulting from the approximation employed
  in the main part of the paper (limiting single traces to length $m$
  but otherwise ignoring potential trace relations).}
\label{fig-nf}
\end{figure}

Fig.~\ref{fig-nf} shows a comparison of ${\cal N}$ and ${\cal N}_{\rm
  ap}$ for some sample values of parameters.  We were able to
compute\footnote{We thank M.~Marino for CHAR, a Mathematica routine
  computing characters of symmetric groups.}  both expressions for all
values of $m$ in the range $K\leq 100$, $F\leq 7$.  We found that the
agreement is very good.  In all examples, our approximation was either
exactly correct (${\cal N}_{\rm ap}(F)={\cal N}(F)$) or slightly
smaller than the true number of states.  This is consistent with our
expectation that a small number of single traces of length greater
than $m$ do contribute to independent states (but they are discarded
in our approximation, giving a small undercount), while the trace
relations ignored in our approximation eliminate only a tiny or
vanishing number of states (giving a tiny overcount).

In concluding this appendix, two more remarks are in order.  First, it
is possible that both the under- and overcount are quite large but
happen to cancel almost precisely.  In this case our approximation
would not be give a reliable set of representations. 

Second, we have computed ${\cal N}$ only in the the regime $F\lesssim
km^2$.  Thus our results are consistent with the earlier thermodynamic
argument suggesting that trace relations should be negligible below
$F\sim km^2$.  However, we were unable so far to compute ${\cal N}(F)$
for any $F>7$, so our numerical work alone cannot pin down the onset
of trace relations (which could be already around $F\sim km$, for
example, without being diagnosed by our numerical results).  It seems
likely that our method of computing ${\cal N}(F)$ can be significantly
improved to achieve higher speeds and larger $F$.

\section{Proofs}

This Appendix contains proofs of two statements made in the main text.

\subsection{Distinguishable particles}
\label{sec-dist}

Here we prove the inclusion relation Eq.~(\ref{eq-qu}) for a pair of
toy spectra in which traces and particles are treated as
distinguishable.  That is, we drop the outer symmetrization
requirement in Eqs.~(\ref{eq-jkl}) and (\ref{eq-klm}).  The
corresponding spectra are
\begin{eqnarray}
  T''(n^m)_F &=& \bigoplus_{P(F\backslash m) = (q_1^{p_1}\cdots q_l^{p_l})}~
  \prod_{i=1}^l (\mathsf{sym}~R(n)_1^{q_i})^{p_i}  ~;
\label{eq-jklm}\\
  T^{\rm prod}(n^m)_F &=& \bigoplus_{P(F/m) = (q_1^{p_1}\cdots q_l^{p_l})}~ 
  \prod_{i=1}^l (\mathsf{sym}~R(n)_1^{q_i})^{p_i}~.
\label{eq-klmn}
\end{eqnarray}
We will now show that $T''(n^m)_F\supseteq T^{\rm prod}(n^m)_F$.

The auxiliary Young diagrams for these toy spectra are unaffected by
this modification.  Its only effect is to change the product of rows
into a regular, rather than a symmetrized product, when the
representations are computed.

We have already learned that auxiliary diagrams appearing in both
spectra can be canceled.  At every $m$ and $F$, the remaining
diagrams will be related to each other by conjugation.  All uncancelled
diagrams in the full spectrum will be higher than wide, and
vice-versa for the product spectrum.  

One can then deform each remaining auxiliary diagram of the full
spectrum into its conjugate diagram (corresponding to one of the
remaining auxiliary diagrams of the product spectrum) by a sequence of
steps which strictly reduce the number of representations generated by
the auxiliary diagram.  

Each step takes one box and moves it to a higher row.  Because there
is no need to symmetrize rows, the effect of this move will be limited
to the product between the origin row and the target row.  Suppose the
length of the origin row before the move is $l$ and the length of the
target row is $k$.  By construction, $k\geq l$.  Originally the two
rows contribute
\begin{equation}
(k) \cdot (l) = (k+l) \oplus (k+l-1~~1) \oplus \cdots 
\oplus (k+1~~l-1) \oplus (k~~l)
\label{eq-trade1}
\end{equation}
to the product of rows.  After moving the box, the contribution will
be
\begin{equation}
(k+1) \cdot (l-1) = (k+l) \oplus (k+l-1~~1) \oplus \cdots 
\oplus (k+1~~l-1)~.
\label{eq-trade2}
\end{equation}
The latter is a subset of the former since the $(k l)$ representation
is missing.  By a sequence of such steps one arrives at the conjugate
of the original Young diagram.  

Thus each uncancelled auxiliary diagram in the full spectrum leads to
representations which are a superset of those generated by the
corresponding (conjugate) leftover diagram in the product spectrum.

\subsection{Mapping states with $U(1)$ spacetime symmetry}
\label{sec-comproof}

Here we prove a result constraining the mapping between states with no
center-of-mass excitation (see Sec.~\ref{sec-com}) in a toy theory
with symmetry group $U(1)$ (see Sec.~\ref{sec-conjugate} for its
simplifying properties, which allow an exact matching of spectra).
Recall that the c.o.m.\ ground states in the full spectrum are in
one-to-one correspondence with auxiliary Young diagrams whose first
and second column are equally high. (This is true whether we think of
this diagram as controlling traces of powers of $B$, or of $A$, as
defined in Sec.~\ref{sec-com}.)

A {\em blunt diagram\/} is a Young diagram whose first and second row
have equal length.  Blunt diagrams are conjugate to the 
c.o.m.\ ground state diagrams in the full spectrum, and so we expect
that they will play a special role in characterizing c.o.m.\ ground
states of the product spectrum.

We prove that a basis of c.o.m.\ ground states in the product spectrum
can be chosen so that each basis element is a linear combination
containing precisely one blunt auxiliary diagram.

To prove this, let us first understand how the c.o.m.\ annihilation
operator acts on an arbitrary diagram.  Here is the list of diagrams
with $F=5$, which get converted into sums of $F=4$ states.
\begin{eqnarray}
\tableau{5} & \to & 5 \tableau{4}^*  \\
\tableau{4 1} & \to & \tableau{4} \oplus 4 \tableau{3 1}^* \\
\tableau{3 2} & \to & 2 \tableau{3 1} \oplus 3 \tableau{2 2}^* \\
\tableau{3 1 1} & \to & \tableau{3 1} \oplus 3 \tableau{2 1 1}^* \\
\tableau{2 2 1}~! & \to & \tableau{2 2} \oplus 4 \tableau{2 1 1} \\
\tableau{2 1 1 1} & \to & \tableau{2 1 1} \oplus 2 \tableau{1 1 1 1}^* \\
\tableau{1 1 1 1 1}~! & \to & \tableau{1 1 1 1}
\end{eqnarray}
On the left are the $F=5$ ``parent'' diagrams on which $a_{\rm com}$
acts; on the right are $F=4$ daughters.  Note that the sums here are
linear superpositions of states, not an addition of representations
($\oplus$).

One finds these relations by working explicitly with operators, and it
is straightforward to generalize them.  The general rule is that the
daughters are obtained by all legal ways to remove one box from a
given parent.  Each such way corresponds to one ``block'' in the
parent diagram formed by rows of equal length.  The coefficients of
daughters are given by the number of boxes in the corresponding block
on the left.

In the above list, we have marked the blunt parent diagrams by
exclamation marks.  On the right, we have added stars to certain
daughter diagrams which have a special property that will be useful to
us: each can be reverted back into its $F=5$ parent diagram by adding
a box on the top right.  Obviously none of the daughters of blunt
parents have this property, and exactly one daughter of any non-blunt
parents does.

To show that any blunt diagram can be combined with non-blunt diagrams
to form a c.o.m.\ ground state, we need to cancel off the daughters.
We give a general argument but it may be instructive to follow along
using the above example.  Add one box to the top right of each
daughter of the blunt diagram.  This will give us a non-blunt diagram
that can be used to cancel every daughter of the blunt diagram, at the
cost of introducing new daughters.  However, none of these new
daughters will carry a star (since the starred daughters were used to
cancel the blunt diagram's daughters).  Hence one can continue adding
a box to the top right and canceling.  This procedure always
increases the number of columns but keeps the number of boxes fixed;
hence it terminates.  This completes the first part of our proof.

For the second part, we must show that a product state with no c.o.m.\
excitation cannot be obtained purely from non-blunt diagrams.  Suppose
the opposite and start with any non-blunt diagram.  One of its
daughters is starred and can only be canceled by daughters of another
diagram obtained by adding a box to some row other than the first row.
By assumption this diagram is non-blunt and hence will produce another
starred daughter.  Continuing in this manner, we never increase the
length of the first row, and hence we must eventually be increasing
the number of rows.  This will eventually require a number of rows
larger than $m$ which is not allowed, or for $m\geq F$ it will force
us to use the one-column parent diagram, which is blunt.  This
contradicts our assumption, completing the proof.

\bibliographystyle{board}
\bibliography{all}

\begin{thebibliography}{10}

\bibitem{BFSS}
T.~Banks, W.~Fischler, S.~H. Shenker and L.~Susskind: {\em {M} theory as a
  matrix model: {A} conjecture\/}. Phys. Rev. D {\bf 55}, 5112 (1997),
  hep-th/9610043.

\bibitem{Mal97}
J.~Maldacena: {\em The large {$N$} limit of superconformal field theories and
  supergravity\/}. Adv. Theor. Math. Phys. {\bf 2}, 231 (1998), hep-th/9711200.

\bibitem{BMN}
D.~Berenstein, J.~M. Maldacena and H.~Nastase: {\em Strings in flat space and
  pp waves from {N} = 4 super {Y}ang {M}ills\/}. JHEP {\bf 04}, 013 (2002),
  hep-th/0202021.

\bibitem{Tay01}
W.~Taylor: {\em M(atrix) theory: {M}atrix quantum mechanics as a fundamental
  theory\/}. Rev. Mod. Phys. {\bf 73}, 419 (2001), hep-th/0101126.

\bibitem{DSV1}
K.~Dasgupta, M.~M. Sheikh-Jabbari and M.~Van~Raamsdonk: {\em Matrix
  perturbation theory for {M}-theory on a {PP}-wave\/}. JHEP {\bf 05}, 056
  (2002), hep-th/0205185.

\bibitem{DSV2}
K.~Dasgupta, M.~M. Sheikh-Jabbari and M.~Van~Raamsdonk: {\em Protected
  multiplets of {M}-theory on a plane wave\/}. JHEP {\bf 09}, 021 (2002),
  hep-th/0207050.

\bibitem{M5}
J.~Maldacena, M.~M. Sheikh-Jabbari and M.~Van~Raamsdonk: {\em Transverse
  fivebranes in matrix theory\/}. JHEP {\bf 01}, 038 (2003), hep-th/0211139.

\bibitem{Sus97}
L.~Susskind: {\em Another conjecture about {M}(atrix) theory\/}  (1997),
  hep-th/9704080.

\bibitem{BahBar82}
A.~Baha~Balantekin and I.~Bars: {\em Branching rules for the supergroup
  {SU(N/M)} from those of {SU(N+M)}\/}. J. Math. Phys. {\bf 23}, 1239 (1982).

\bibitem{AhaSon96}
O.~Aharony, J.~Sonnenschein and S.~Yankielowicz: {\em Interactions of strings
  and {D}-branes from {M} theory\/}. Nucl. Phys. {\bf B474}, 309 (1996),
  hep-th/9603009.

\bibitem{LiE}
{\em {LiE}: A computer algebra package for {L}ie group computations\/}, \\
  \verb+http://wwwmathlabo.univ-poitiers.fr/~maavl/LiE/+~~.

\bibitem{FauJar05}
B.~Fauser, P.~Jarvis, R.~King and B.~Wybourne: {\em New branching rules induced
  by plethysm\/}  (2005), math-ph/0505037.

\bibitem{VanU}
M.~Van~Raamsdonk: {\em unpublished\/}.

\bibitem{AhaMar03}
O.~Aharony, J.~Marsano, S.~Minwalla, K.~Papadodimas and M.~Van~Raamsdonk: {\em
  The {H}agedorn / deconfinement phase transition in weakly coupled large {N}
  gauge theories\/}. Adv. Theor. Math. Phys. {\bf 8}, 603 (2004),
  hep-th/0310285.

\bibitem{AgaOog04}
M.~Aganagic, H.~Ooguri, N.~Saulina and C.~Vafa: {\em Black holes, q-deformed 2d
  {Y}ang-{M}ills, and non-perturbative topological strings\/}. Nucl. Phys. {\bf
  B715}, 304 (2005), hep-th/0411280.

\end{thebibliography}
\end{document}